\documentclass{aa}
\usepackage{txfonts}
\usepackage{footnote}
\usepackage{graphics}
\usepackage{epsfig}
\usepackage{natbib}
\usepackage{lscape}
\usepackage[usenames,dvipsnames,table]{xcolor}
\usepackage{mdwlist}
\usepackage{rotating}
\usepackage{array,longtable}

\usepackage[totalwidth = 17.8cm, totalheight=23.0cm]{geometry}

\begin{document}
\title{The AGB population in IC 1613 using JHK photometry}

\author{L. F. Sibbons\inst{1}
        \and 
        S. G. Ryan\inst{1}
        \and 
        M. Irwin\inst{2}
        \and
        R. Napiwotzki\inst{1}
        }

\offprints{L.Sibbons1@herts.ac.uk}

\institute{School of Physics, Astronomy and Mathematics, University of
  Hertfordshire, College Lane, Hatfield AL10 9AB, United Kingdom
 \and
  University of Cambridge, Institute of Astronomy, Madingley Rd, Cambridge CB3 OHA, United Kingdom}

\date{Received 18 Apr 2014/ Accepted 12 Sept 2014}

\titlerunning{AGB Population of IC1613}
 
\authorrunning{Sibbons et al}

\abstract{A member of the Local Group, IC~1613 is a gas rich irregular
  dwarf galaxy that appears to have formed stars continuously over the
  last $10$ Gyr and is relatively independent of external influences
  from other galaxies.}{This paper aims to study the spatial
  distribution of the asymptotic giant branch (AGB) population in IC~1613 and its
  metallicity.}{Using WFCAM on UKIRT, high quality $JHK$ photometry of
  an area of $0.8$~deg$^{2}$ centered on IC~1613 was obtained. The
  data have been used to isolate the C- and M-type components of the
  AGB population and using their number ratio, C/M, a global mean
  metallicity has been derived. The metallicity and the
  TRGB magnitude have been studied as a function of distance from the
  galactic centre and as a function of azimuthal angle.}{The tip of
  the RGB (TRGB) has been found at $K_{0} = 18.25 \pm 0.15$~mag. The colour
  separation between the C- and M-type components of the AGB
  population has been located at ($J-K$)~$= 1.15 \pm 0.05$~mag, giving
  a global C/M ratio of $0.52 \pm 0.04$ and from this an iron
  abundance of [Fe/H]~$= -1.26 \pm 0.07$~dex has been calculated.}{The
  AGB population has been detected out to a radial distance of
  $4.5$~kpc in the de-projected plane of the galaxy. The measured TRGB
  is consistent with previous measurements and no significant
  variation is detected in the TRGB or in metallicity either with
  galactocentric distance or azimuthal angle.}\thanks{Tables 1 and 2 are only available in electronic form at the CDS via anonymous ftp to cdsarc.u-strasbg.fr (130.79.128.5) or via http://cdsweb.u-strasbg.fr/cgi-bin/qcat?J/A+A/}


\maketitle

\section{Introduction}
\label{intro}
The dwarf galaxy IC~1613 is frequently described as being Magellanic like or a
`typical' dwarf irregular (dIrr). With a luminosity of $M_{V} = -14.9$~mag it is ranked $6^{th}$ in brightness among the dIrr's of the
Local Group (LG) \citep{2009A&A...493.1075B}. At a distance of $\sim
500$~kpc from M31, IC~1613 is more closely associated with that galaxy than the Milky
Way (MW), although there is some disagreement in the literature as to
whether it is a satellite of M31 \citep{1989AJ.....98.1274L,2006MNRAS.365..902M,2007A&A...466..875B}. Estimates
of the distance to IC~1613 from the MW have been made using several different
distance-luminosity indicators including Cepheids $(m-M)_{0} = 24.291
\pm 0.035$ ($721$~kpc) \citep{2006ApJ...642..216P}, RR Lyraes
$(m-M)_{0} = 24.10 \pm 0.27$ ($660$~kpc) \citep{1992AJ....104.1072S},
the tip of the red giant branch (TRGB) $(m-M)_{0} = 24.29
\pm 0.12$ ($721$~kpc) \citep{1999AJ....118.1657C} and the red clump
$(m-M)_{0} = 24.30 \pm 0.09$ ($724$~kpc)
\citep{2001ApJ...550..554D}. \citet{2010ApJ...712.1259B} presented an
average distance modulus of $(m-M)_{0} = 24.40 \pm 0.014$  ($758$~kpc)
based on their own derived value and those in the literature. We adopt
this average value for our work, meaning $1'$ equates to
$0.22$~kpc.

IC~1613 is a gas rich galaxy that appears to have experienced
star formation over most of the life of the Universe as evidenced by
the presence of an old ($>10$~Gyr) RR Lyrae population
\citep{1992AJ....104.1072S,2001ApJ...550..554D,2010ApJ...712.1259B},
an intermediate-age ($1-10$~Gyr) red giant branch (RGB) and asymptotic
branch (AGB) population
\citep{1988AJ.....96.1248F,2009JASS...26..421J} and a younger ($< 1$~Gyr) blue population of stars
\citep{1991ApJ...369..372H,2007AJ....134.2318T}. At present there
seems to be only one recent and ongoing region of star formation in
the North East (NE) of the galaxy, but the hole in the H~I gas in the
South East (SE) suggests that there has been star formation in that
region in the recent past
\citep{2002A&AT...21..223L,1989AJ.....98.1274L}. 

While star formation in IC~1613 has been effectively continuous the stellar population is dominated by stars formed at intermediate times \citep{1999AJ....118.1657C}. This is the result of variations in the rate of star formation with time and location, at least within the $~ 7'$ of the dynamic center of the galaxy \citep{2003ApJ...596..253S}. Using optical photometry of a small field near the center of IC~1613, taken with the Wide-Field Planetary Camera 2 (WFPC2), \citet{1999AJ....118.1657C} estimated that the star formations rate (SFR) had declined by $\sim 50\%$ over the last $400-900$~Myr and is currently $\sim 3.5 \times 10^{-4}$~M$_{\odot}$~yr$^{-1}$. \citet{2003ApJ...596..253S}, who compared WFPC2 photometry of a second field approximately $6'.7$ South West of that observed by \citet{1999AJ....118.1657C} with colour-magnitude diagrams (CMDs) simulated from theoretical stellar models, also concluded that the SFR in IC~1613 had declined in the last Gyr; following a period of enhanced star formation between $3$ and $6$~Gyr ago. However, \citet{2003ApJ...596..253S} found that the decline in the SFR differed between the inner and outer fields, with the SFR in the central field observed by \citet{1999AJ....118.1657C} maintaining a higher average level. More recently \citet{2014ApJ...786...44S} used the HST Advanced Camera for Surveys (ACS) to obtain very deep ($I \sim 29$~mag) optical photometry of another field between the core and half-light radius of IC~1613. The depth of these observations means that they can reach a time resolution of $\sim 1$~Gyr at the oldest ages and \citet{2014ApJ...786...44S} argue that the position of the observed field means that it can be treated as being representative of the global star formation history (SFH). They find that star formation has been nearly continuous, with an almost constant SFR during the first $6$~Gyr and varying by, at most, a factor of two over the entire lifetime of the galaxy.

In addition to deep photometric studies of isolated fields 
near the centre of the galaxy, where the younger stars are concentrated,
several studies have been made of 
the distribution and metallicity of the different stellar
populations across IC~1613 in order to better understand its SFH. 
\citet{2000AJ....119.2780A} detected C-type stars out to a
radius of $15'$, while \citet{2007A&A...466..875B} traced the giant
population out to $23'$, finding though that the young main sequence
population is restricted to the central portion of the galaxy ($\leq
10'$). The metallicity of IC1613 is often found to be comparable to the Small
Magellanic Cloud or slightly more metal poor \citep{1999AJ....118.1657C}. Estimates of the iron abundance depend on 
the stellar population under investigation. Using optical
photometry, \citet{2003ApJ...596..253S} found that [Fe/H] has increased from $-1.30$ at the earliest
times to $-0.70$~dex today. The same authors quoted abundances of
$-0.80 \pm 0.20$~dex and $-1.07$~dex for H~II regions in the galaxy
based on measurements of the nebular oxygen abundance made by
\citet{1989ApJ...347..875S} and \citet{2003A&A...401..141L}
respectively. This is consistent with the findings of
\citet{2007AJ....134.2318T} who estimated the overall metallicity of
the young stellar population to be [Fe/H]~$= -0.67 \pm 0.09$~dex from
the spectra of $3$ M-type supergiants and with those of
\citet{1999AJ....118.1657C} who estimated the [Fe/H] abundance of the
blue and red supergiants to be $\leq -1.0$~dex. Multiple [Fe/H]
measurements have also been made of the intermediate- and old-age
populations. Based on the average ($V - I$) colour of the population at
the TRGB, \citet{1988AJ.....96.1248F}
and \citet{1999AJ....118.1657C} estimated the metallicity of the RGB
population to be [Fe/H]~$ =-1.30$~dex and [Fe/H]~$= -1.40 \pm 0.30$~dex respectively. These measurements are in good agreement with those
of \citet{2004oee..sympE..62Z}, who find a mean abundance of [Fe/H]~$=
-1.30$~dex from the equivalent width of two Ca~II triplet lines in the
spectra of a sample of RGB stars. \citet{2011AJ....141..194G}
estimated the metallicity of the older RGB population to be [Fe/H]~$=
-1.50 \pm 0.08$~dex based on their optical colours, while
\citet{2001ApJ...550..554D} quote a more metal rich value of
[Fe/H]~$= -1.15 \pm 0.20$~dex based on the interpolation of the
isochrones by \citet{2000A&AS..141..371G}. For the RR Lyrae population
($> 10$~Gyr) \citet{2001ApJ...550..554D} estimate a mean metallicity
of [Fe/H]~$= -1.30 \pm 0.2$~dex. This is more metal rich than other
estimates by  \citet{2002A&A...394...33T} and
\citet{1999AJ....118.1657C} who estimate the metallicity of the old
population to be [Fe/H]~$= -1.75 \pm 0.20$ and [Fe/H]~$= -1.80$ to
$-2.0$~dex respectively, based on optical photometry of the giant
population. We will examine the intermediate age ($1-10$~Gyr) AGB
population and derive its metallicity using $JHK$
photometry and the C/M ratio. 

During the thermally pulsing AGB (TP-AGB) phase, mixing
mechanisms can dredge up triple-$\alpha$ processed material from the
He-burning shell to the stellar atmosphere. The dredged up material
can cause the dominant metal to change from oxygen to carbon. Sources
which have more oxygen than carbon (C/O $<1$) in their atmospheres are
known as oxygen-rich or M-type, while those that have more carbon than
oxygen (C/M $>1$) are known as carbon-rich or C-type. The ratio
between the number of C- and the number of M-type stars is known as
the C/M ratio. This ratio is often used as an indirect indicator of
metallicity in the environment in which those stars formed, as at
lower metallicities the transformation from an initially O-rich
atmosphere to a C-rich one is easier as fewer dredge-up events are
required \citep{1983ARA&A..21..271I,1978Natur.271..638B}.

The
structure of this paper is as follows. In Sect. \ref{obs} we present
our data and the selection of sources for our analysis, in
Sect. \ref{anal} \& \ref{results} we analyse the data and present out
results, followed by a discussion and our conclusions in
Sect. \ref{finish} \& \ref{cons}.

\section{Observations and data reduction}
\label{obs}

\subsection{$JHK$ photometry}
\label{phot}
Images of IC~1613, centered on $\alpha = 01^{h}04^{m}54^{s}, \delta =
02^{\circ}07'57.4''$ were obtained as part of a larger project to
survey the AGB population of Local Group galaxies (PI Irwin). Using
the Wide Field CAMera (WFCAM) mounted on UKIRT in Hawai'i on the night
of the 5 August 2008, images were obtained in the near-infrared (NIR)
bands $J, H$ and $K$ covering an area of $0.80$~deg$^{2}$ on the
sky. The total exposure time per pixel for each band was $200$s from
the co-addition of $4$ frames each taken using a dithered pattern of
$5$ positions with an exposure time of $10$s in each position. The
average seeing on the night of the observations was $0.64''$.

The
data are similar to those used by \citet{2012A&A...540A.135S} to
investigate the AGB population of the LG dwarf NGC 6822. Details of
the reduction and calibration of the photometry using the 2MASS point
source catalogue are as given by \citet{2012A&A...540A.135S}. The magnitudes and colours given here are on the
WFCAM instrumental system (see \citealt{2009MNRAS.394..675H} for
transformations) and have been corrected for foreground reddening in
the direction of IC~1613 using the extinction map of
\citet{1998ApJ...500..525S}, which gives E(B-V)~$= 0.02-0.03$~mag. Internal obscuration in IC~1613 is also low, estimated to be
between E(B-V)~$= 0.06 \pm 0.02$ and $0.09 \pm 0.019$
\citep{1999A&AS..134...21G,2006ApJ...642..216P} in the dusty star
forming regions of the galaxy. No correction has been made to account
for internal reddening in the galaxy but as the effects of reddening
are reduced in the NIR (E(J-K)~$= 0.010 - 0.015$~mag) the impact on
our results is expected to be negligible. All magnitudes and colours
are presented in their dereddened (foreground only) form. 

\begin{figure} 
\includegraphics[scale = 0.4]{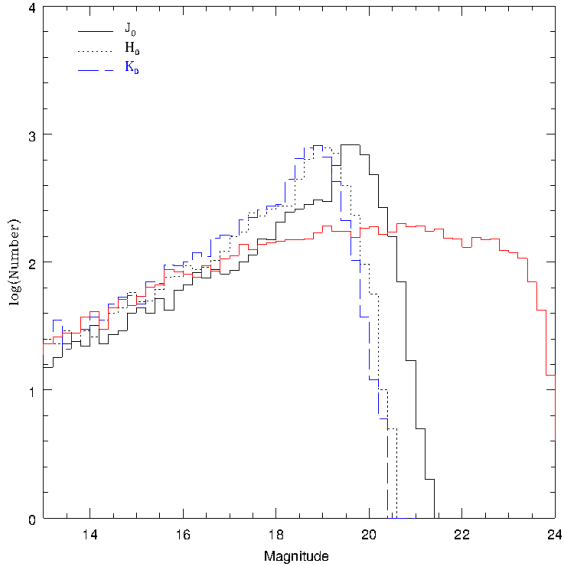}
\caption{\tiny{Log Number vs. linear magnitude in $0.2$~mag bins. The 
red line shows the magnitude distribution of the TRILEGAL synthesised foreground 
population in the $K$-band.}} 
\label{comp}
\end{figure}

In Fig. \ref{comp} we present the magnitude distribution of our
sources in all three bands and a synthetic foreground generated (in
the $K$-band) using the TRILEGAL code \citep{2005A&A...436..895G}. The
real and the synthetic data have similar bright-star distributions, until $\sim
17-18$~mag in $J-,H-$ and $K$-bands where they begin to diverge. The
synthetic population (which is assumed to be complete) becomes flatter
at $K \gtrsim 18$~mag before starting to gradually decline at $K \sim
22$~mag. The number of sources we detect however steadily increases fainter than $K
\sim 17-18$~mag, where AGB stars in IC~1613 are detected, before
showing a distinct discontinuity at $J \sim 19.20$~mag, $H \sim 18.40$~mag and $K \sim 18.20$~mag reflecting a further increase in the number of
sources at the TRGB. The presence of giants in IC~1613 is more obvious
than it was in the case of NGC 6822 \citep[their
Fig. 3]{2012A&A...540A.135S} due to the much lower foreground
contamination in the direction of IC~1613 (\textit{l}~$=
129^{\circ}.74$, \textit{b}~$= -60^{\circ}.58$) compared to NGC 6822
(\textit{l} $= 25^{\circ}.34$, \textit{b} $= -18^{\circ}.39$). These approximate TRGB values are supported by the published values by
\citet{2011AJ....141..194G} ($J_{TRGB} = 19.19 \pm  0.08$~mag,
$K_{TRGB} = 18.13 \pm 0.08$~mag) and by  \citet{2009JASS...26..421J}
($J_{TRGB} = 19.10$~mag, $H_{TRGB} = 18.40$~mag, $K_{s_{TRGB}} =
18.0$~mag). Our estimate of the TRGB magnitude is refined in
Sect. \ref{trgb}.

The completeness of our data in each band was also
inferred from Fig. \ref{comp}. As noted above, the synthetic
distribution begins to decline at $K \sim 22.0$~mag; this decline is
attributed to a change in the MW population - possibly the edge of the
disk. For our $JHK$ data we see a sharp decline at much brighter
magnitudes, which we attribute to the declining completeness of the
data. The data are assumed to be complete until the peak of the
distribution in each band, $19.7, 19.3$ and $18.9$~mag in $J,H$ and
$K$ respectively. Beyond this, we approximate the completeness level by
normalising to the observed star count at the peak of each band using this computational convenience. The
completeness level in each band falls to the $50\%$ level at around
$20.6$~mag in $J$, $19.8$~mag in $H$ and $19.4$~mag in $K$. The data
are therefore sufficiently complete for our intended study of the AGB
population based on the estimated position of the TRGB discussed
above.

\subsection{Source selection}
\label{flags}
From the raw images, a catalogue of $\sim 27,700$ sources was
derived. Each source was assigned a flag depending on the quality of
the data in each photometric band. Sources were flagged as: saturated,
noise-like, non-stellar, probably-stellar, stellar, compact
non-stellar or poor astrometry match. In our analysis of a similar
data set for NGC 6822 we selected only those sources flagged as
stellar or probably-stellar in all three bands. However, spectroscopic
follow up \citep[in preparation]{2013..InPreparation...S} has led us
to relax this criterion as it results in the rejection of many
genuine stellar sources. If only those sources meeting the stringent
three flag criterion are selected, the number of sources in our primary
photometric sample is reduced to $4858$. If the more relaxed two flag
criterion is applied, in which we require only two bands to have a
stellar or probably-stellar flag, but still require a magnitude
measurement in all three photometric bands, a photometric sample of
$7008$ sources is obtained.

By selecting sources that have been classified as stellar or probably stellar in at least two bands, we ensure that we obtain a larger sample, but are still able to exclude potential contamination from background galaxies, which are likely to have a non-stellar classification. The potential extent of the contamination from these background sources is demonstrated in Fig. \ref{flags} where we present two ($J-K$, $K$) CMDs. The first of these shows a sample of sources that have been classified as non-stellar in both the $J$- and $K$-bands, while the second shows those sources that have been retained using the two band and three band flag criterion used to select our sample. If our two flag criterion was relaxed further (e.g. to only one flag) the potential contamination from non-stellar sources, particularly at redder ($J-K$) colours, would  be significant.

\begin{figure*} 
\includegraphics[scale = 0.45]{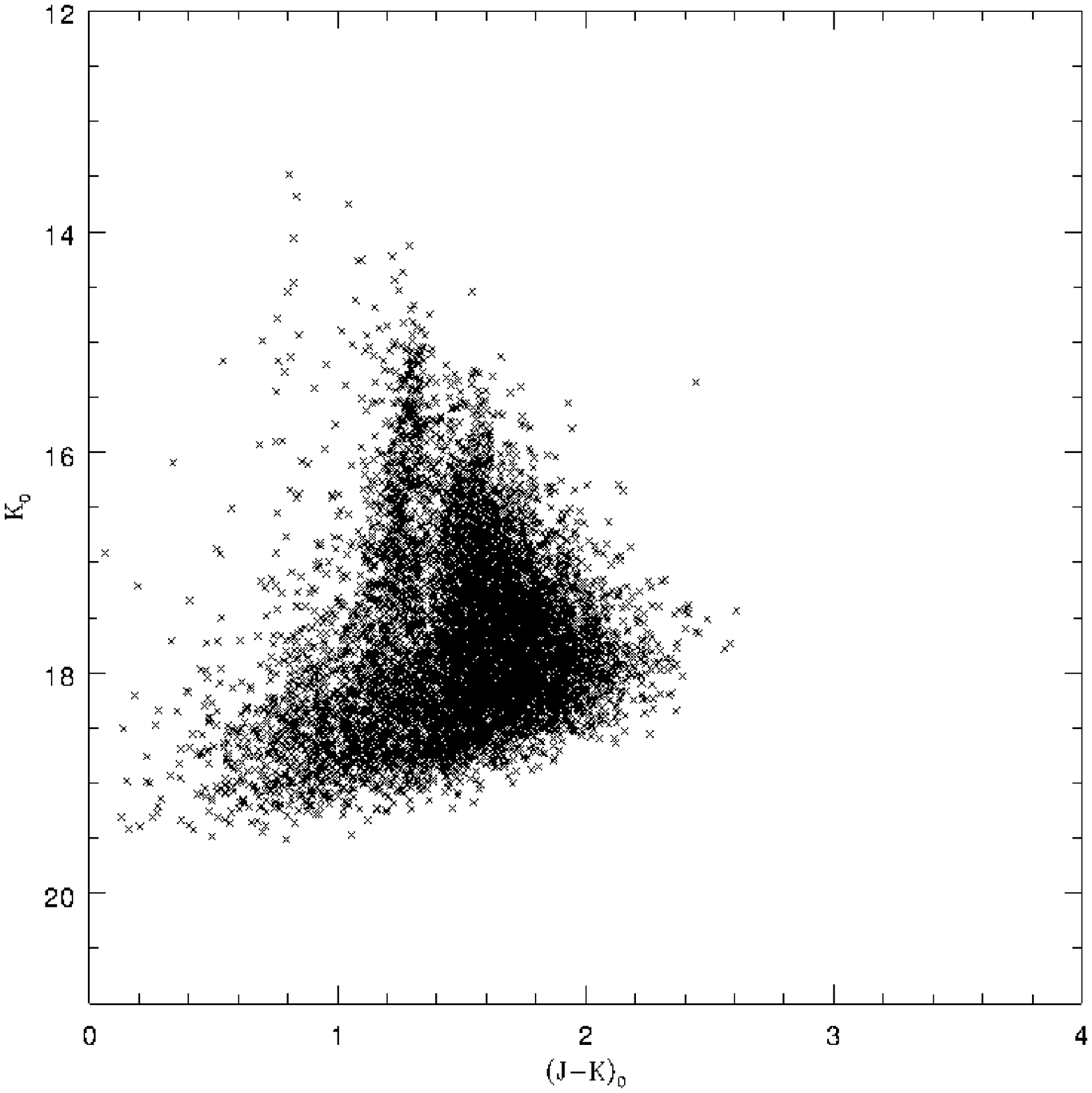}
\includegraphics[scale = 0.45]{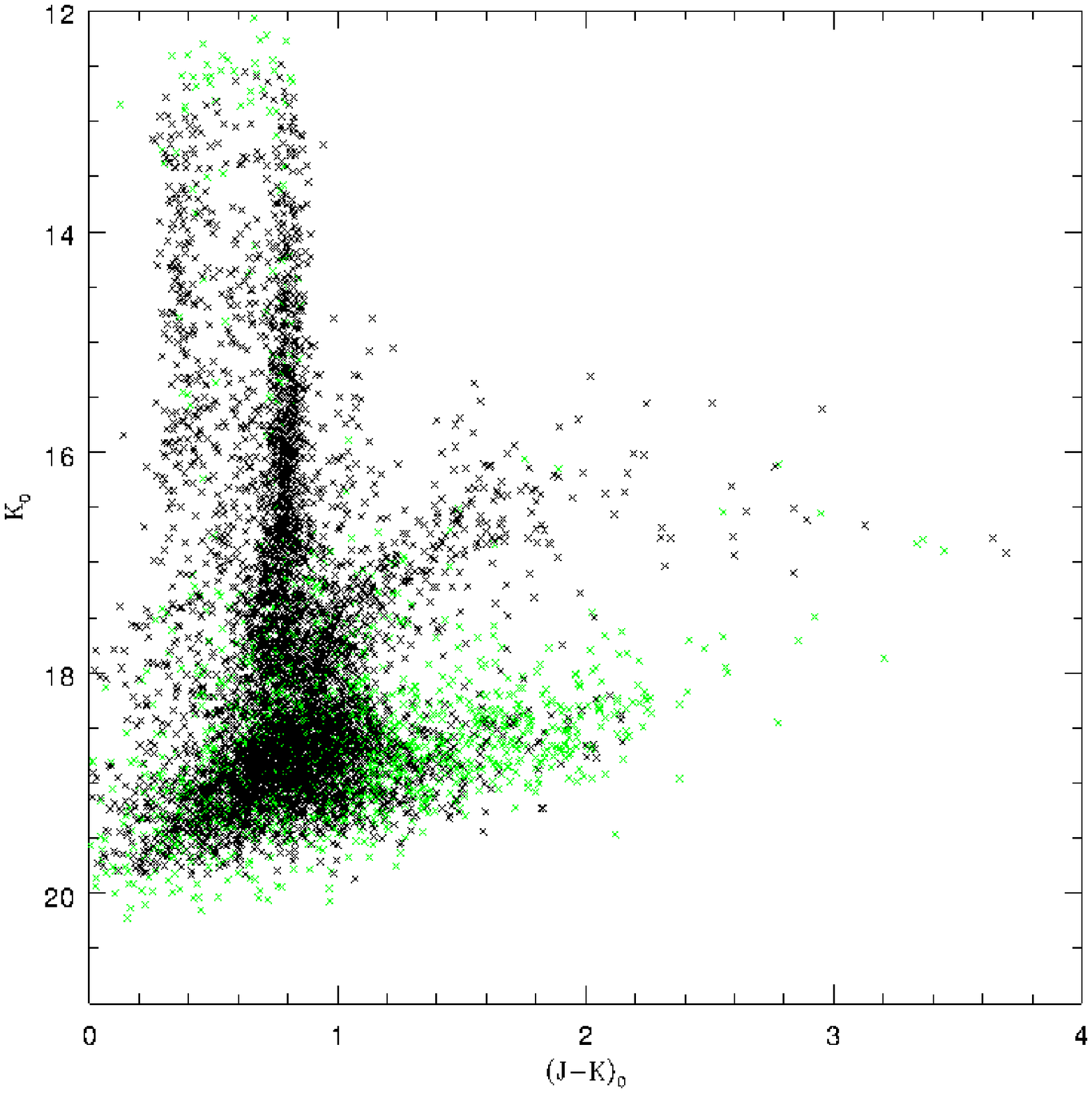}
\caption{\tiny{Left: CMD of those sources with a non-stellar classification in both the $J$- and $K$-bands. Right: CMD of those sources that have been selected using the flag criteria discussed in the text. Sources in black are those that meet the three band flag criteria. Sources in green are those that meet the two band flag criteria.}} 
\label{flags}
\end{figure*}

\begin{figure} 
\includegraphics[scale = 0.4]{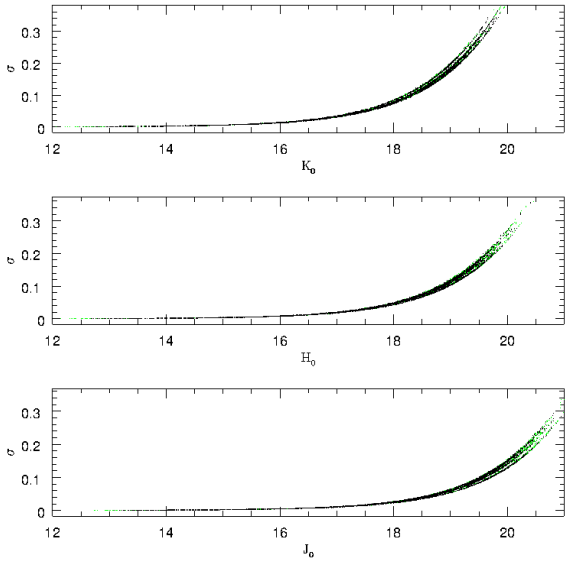}
\caption{\tiny{Error vs. magnitude for each photometric band. In 
black are sources that have been classified as stellar or 
probably-stellar in all three photometric bands. In green are 
the additional sources that have been classified as stellar or 
probably stellar in only two photometric bands.}} 
\label{error}
\end{figure}

In order to confirm that the additional sources retained using the two
flag criterion are of comparable quality to those retained using the
three flag criteria, we plot the error on the photometric measurement
for each source in each band in Fig. \ref{error}. The additional
sources selected using the two flag criterion show a similar error
distribution to those selected using the three flag criteria and
therefore do not lower the overall quality of our photometric
sample. The increase in error at fainter magnitudes is typical of
photometric measurements. We will therefore apply the two flag
criterion and retain the larger photometric sample for our analysis.

\section{Analysis}
\label{anal}

\subsection{Foreground and K-giant removal}
\label{fore}
The high galactic latitude of IC~1613 means that it does not suffer from high levels
of foreground contamination along the line of sight. Nevertheless, any
foreground stars in our photometric sample must be removed before we
can isolate the AGB population of IC~1613. The schematic diagram of
\citet[Fig. A3]{1988PASP..100.1134B} shows that a ($H-K, J-H$)
colour-colour diagram can be employed to separate K- and M-type dwarf sources from
C- and M-type giants, as they diverge in ($J-H$). To establish the
appropriate selection criteria, relatively uncontaminated samples of
the galactic and foreground populations are needed for comparison. 

\begin{figure} 
\includegraphics[scale = 0.4]{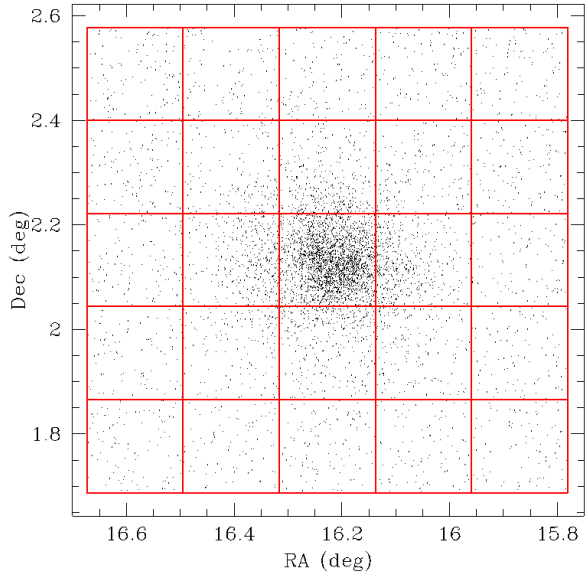}
\caption{\tiny{The observed area has been subdivided into a grid 
of $25$ squares each with dimensions of $\sim 10' \times 10'$. 
The high density of sources in the centre is IC~1613.}} 
\label{grid}
\end{figure}

\begin{figure*} 
\includegraphics[scale = 0.45]{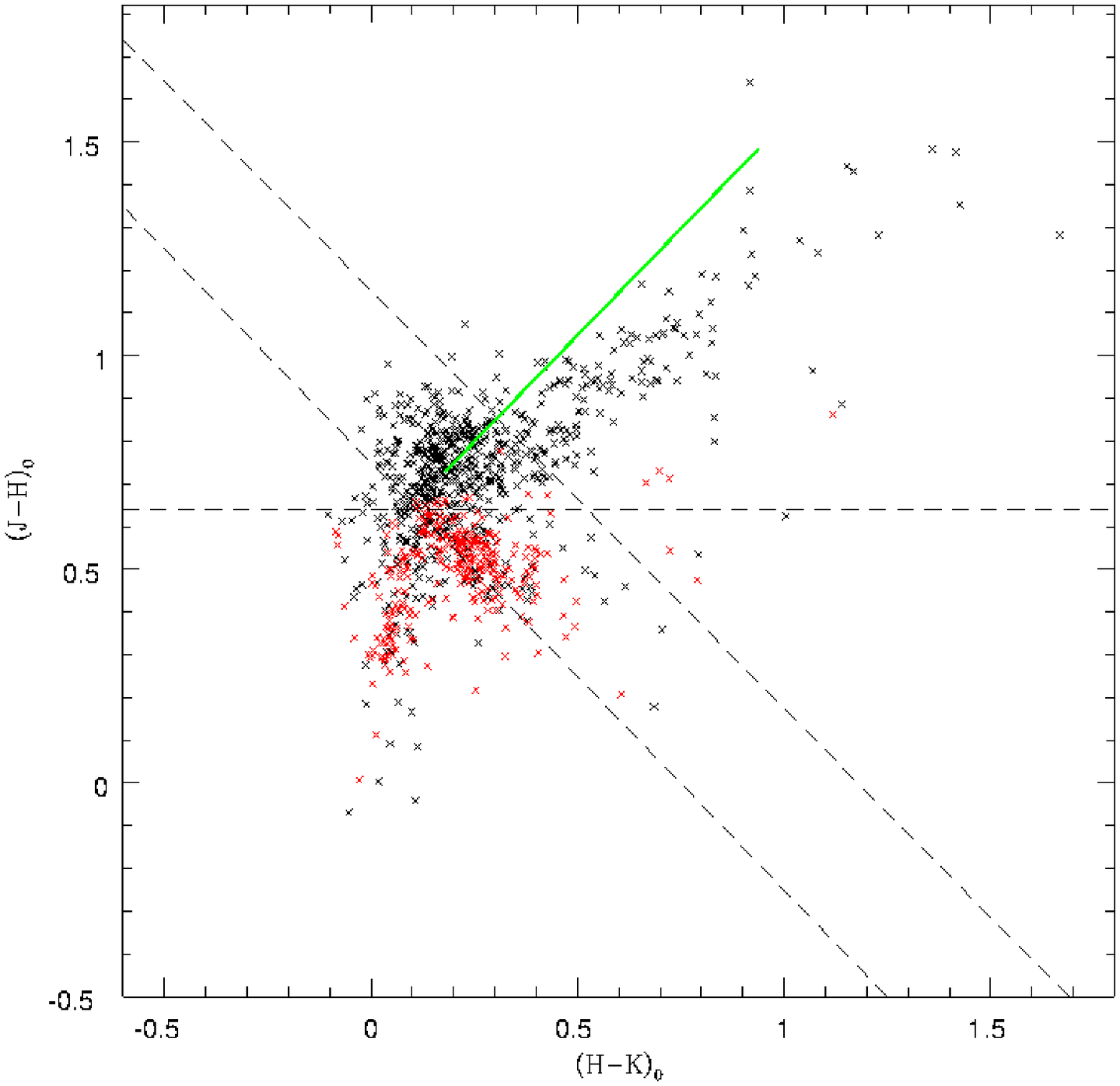}
\includegraphics[scale = 0.45]{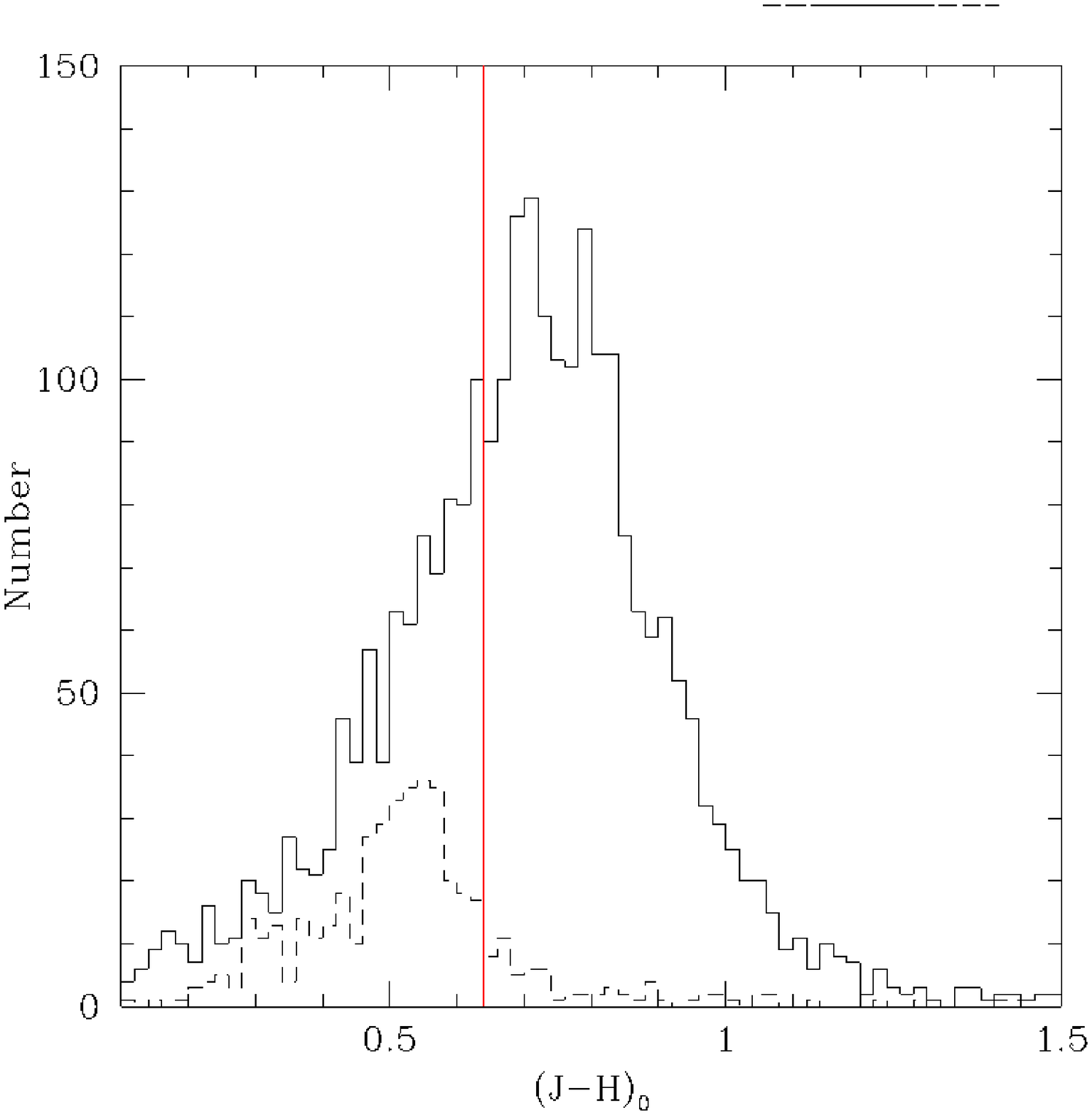}
\caption{\tiny{Left: Colour-colour diagram of sources from the centre 
of the observed area which is dominated by IC~1613 sources (black 
crosses) and sources from the four corner regions of the grid (red crosses) 
where MW sources are expected to dominate. Only sources with $\sigma_{J,H,K} < 0.1$~mag have been shown. The dashed horizontal line
at ($J-H$)~$= 0.64$~mag is the primary foreground removal cutoff
applied to our sample. The dashed diagonal lines at ($J-K$)~$= 0.75$ 
and $1.15$~mag are the colour criteria used for the selection of C- 
and M-type AGB candidates (Sect. \ref{cols}). The solid green line 
shows the trend identified by \citet{2006MNRAS.369..751W} for Galactic and LMC C-type Mira. Right: A ($J-H$) 
colour histogram of the same sources, with bin size $0.02$~mag. The solid line histogram represents 
those sources from the central region, the dashed line represents 
the corner region sources over an area four times larger 
than the central region. The solid red line is at ($J-H$)~$= 0.64$~mag.}} 
\label{CCDfg}
\end{figure*}

To achieve this, the observed area has been subdivided
into smaller regions (Fig. \ref{grid}) of $\sim 10' \times 10'$. Although the central grid square will still be
contaminated by foreground sources, the low density in the direction
of IC~1613 and the high proportion of the galaxy contained in this
part of the grid means that the level of contamination is not likely
to be significant, and the colour distribution of these sources should
primarily reflect that of the IC~1613 giants. In the grid corners the
opposite is expected to be true. Although sources identified as
belonging to IC~1613 have been traced out to a radius of $\sim 23'$
\citep{2007A&A...466..875B}, the corner grid regions lie well
beyond that at $30'$ and the colour distribution of sources in these
regions should give a good indication of the colours of the foreground
sources. It was decided to combine the sources from the four corner squares
as a representative sample of the MW
foreground. A two-colour plot comparing the corner sources with those
in central region can be seen in the left panel of
Fig. \ref{CCDfg}. Only sources with a photometric error ($\sigma_{J,H,K} < 0.1$~mag) 
have been plotted in Fig. \ref{CCDfg} to better show the underlying distribution. 
Although there is some scatter, the distribution of the 
foreground sources mirrors that of the dwarf and K-type sources in the 
\citet{1988PASP..100.1134B} schematic  and there is a clear
separation between the foreground and IC~1613 at ($J-H$)~$\sim 0.65$
mag. A ($J-H$) colour histogram, shown in the right hand panel of Fig. \ref{CCDfg} is used to refine the position of this
colour boundary and shows that the number of sources in the corner grid regions
declines sharply at ($J-H$)~$< 0.64$~mag. This cut is also in good
agreement with the ($J-H$) colours of the TRILEGAL synthesised
foreground, $95\%$ of which have colours bluer than ($J-H$)~$= 0.64$
mag.

A colour cut of ($J-H$)~$= 0.64$~mag is therefore used to remove the bulk
of the foreground sources from our photometric sample. However, as discussed in our previous and forthcoming work
\citep{2012A&A...540A.135S,2013..InPreparation...S} such a colour
criterion is imperfect. As can be seen in the right hand panel of Fig. \ref{CCDfg}, the colour
distribution of the assumed foreground sources extends to redder
($J-H$) colours than the cut off being applied here, and some
foreground contaminants will inevitably be retained. It is also true that
some genuine IC~1613 sources with bluer ($J-H$) colours, in particular 
many K-type giants,  will be removed by the application of this colour 
criterion which we discuss in Sect. \ref{removeFG}. 

\begin{figure} 
\includegraphics[scale = 0.40]{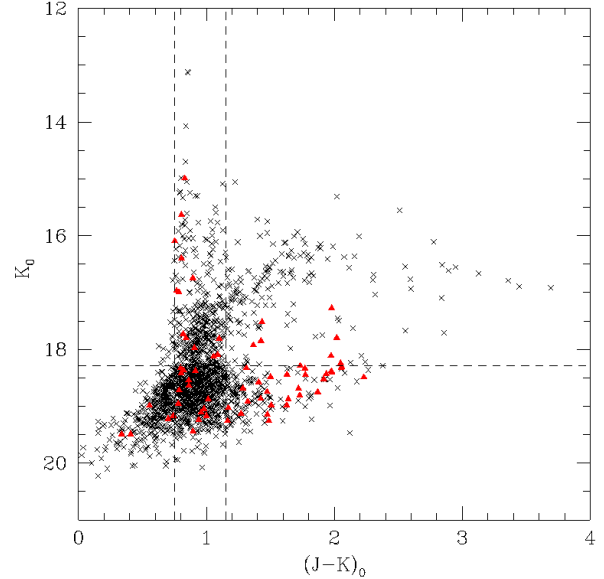}
\caption{\tiny{CMD of sources from the central grid region (black crosses) and from the four corner regions (red triangles) with ($J-H$)~$\geq 0.64$. The dashed horizontal line marks the position of the TRGB at $K = 18.28$~mag (Sect. \ref{trgb}). The dashed vertical lines mark the positions of the blue limit (left) and the boundary between the C- and M-type stars (right) (see Sect. \ref{cols}).}} 
\label{JKfg}
\end{figure}

We can make
an estimate of the remaining foreground contamination. Proceeding
under the assumption that the sources in the outer grid regions with
colours of ($J-H$)~$\geq 0.64$~mag are all foreground sources and that
their redder colours are within the normal range for the foreground
population (TRILEGAL synthesised foreground colours in ($J-H$) can be
as red as $1.36$~mag); although some may be genuine IC~1613 sources,
the number in not likely to be high and we have no way to distinguish
them from the foreground sources at this time. $69$
($15\%$) of the $451$ sources that make up our foreground sample have
colours that are redder than our foreground cut, suggesting the central
grid contains approximately $17$ MW foreground stars after the ($J-H$)
cut has been applied. Fig. \ref{JKfg} shows the CMD of the foreground
`cleaned' central region and those sources from the corner regions
that have colours ($J-H$)~$\geq 0.64$~mag, approximately $70\%$ of which are fainter than the TRGB (Sect. \ref{trgb}).

A second colour cut in ($J-K$) was considered to remove those foreground sources with colours ($J-H$)~$\geq 0.64$~mag. However, any cut in ($J-K$) to remove a significant number of these remaining dwarf sources was found to have a far greater impact on the number of potential M-type giants and it was therefore decided not to proceed with such a cut. We estimate a remaining foreground contamination level of approx. $5$ AGB stars per grid region or $\sim 0.8$ stars per~kpc$^{2}$, brighter than the TRGB, if IC~1613 is inclined at $38^{\circ}$ \citep{1989AJ.....98.1274L}. We expect a residual foreground density of $< 1\%$ among the AGB sources retained in the central region, though this percentage will increase with radial distance from the galactic centre as the stellar density declines.

We have concentrated here on the removal of foreground dwarfs from our 
photometric sample as foreground giant sources are
expected to be so bright that they will have been removed as saturated
sources ($K < 13.55$~mag) by our flag criteria. This is consistent
with the right hand panel of Fig. \ref{flags}, where among the
sources included in our photometric sample as a result of using the
two-band flag criteria we see a trail of very bright ($K < 13.5$~mag)
sources blueward of ($J-K$)~$= 0.80$~mag. An examination of the flag
classifications of these sources shows that with one exception they
have all been classified as saturated in their `non-stellar'
band. However, any impact on our final giant sample will be negligible
as $> 91\%$ of these potential foreground giant sources have been
removed with the application of the blue colour limit at ($J-K$)~$=
0.75$~mag (see Sect. \ref{blue}).  

While our use of the NIR colour-colour diagram has focused on the removal of contaminants from our sample, the same diagram has been used by others to classify AGB and RGB sources as either C- or M-type. Based on the schematic of \citet{1988PASP..100.1134B} the large clump of IC~1613 sources located at ($H-K$)~$\sim 0.15$~mag are likely to be M-type AGB and RGB stars, while the extended tail of sources at redder ($H-K$) and ($J-H$) colours follows the same trend identified \citet{2006MNRAS.369..751W} for Galactic and LMC C-type Miras (Fig. \ref{CCDfg}). During their work on the dwarf spheroidal galaxies Leo~II and Leo~I, \citet{2008MNRAS.388.1185G} and \citet{2010MNRAS.tmp..431H} used this colour-colour diagram to classify the AGB stars in their respective samples as C- or M-type. However, the larger photometric errors in our data compared to those of \citet{2008MNRAS.388.1185G} and \citet{2010MNRAS.tmp..431H}, as well the indistinct colour boundary between the two populations in our sample, mean that we are unable to employ the same technique. Instead we have relied on a ($J-K, K$) CMD to distinguish between the two spectral types as we discuss in Sect. \ref{cols}.

\subsection{Tip of the RGB}
\label{trgb}
The AGB population is isolated using a magnitude cut at the tip
of the red giant branch. The TRGB is one of the  most prominent
features in the magnitude distribution of old- and intermediate-age
populations. The onset of He burning in stars ascending the RGB
results in a striking discontinuity in the magnitude distribution at
the TRGB, as the number of sources at brighter magnitudes (on the AGB)
falls off sharply in comparison with the number of sources at fainter
magnitude (on the RGB) due to the difference in the evolutionary timescales of the two phases.

The position of the TRGB has been located by
applying the Sobel edge detection algorithm
\citep{1993ApJ...417..553L} to a smoothed magnitude distribution using
the modified method of \citet{1996ApJ...461..713S}. The Sobel
algorithm is a first derivative operator that computes the rate of
change (gradient) across an edge, producing a peak where there is a
significant change of slope. Due to the sharp discontinuity in the
magnitude distribution around the TRGB, the largest peak produced by
the filter is expected to be located at the TRGB. A Gaussian is then
fitted to this peak to establish the magnitude and associated error of
the TRGB. However, in order to accurately apply the Sobel filter, it
is necessary that the population being examined is sufficiently large
that there are at least $100$ sources in the range extending one
magnitude fainter than the TRGB \citep{1995AJ....109.1645M,
  2002AJ....124.3222B}. To ensure that this requirement is met, the
Sobel filter is initially applied to the central grid region in
Fig. \ref{grid} as it has the highest density of sources. After the
colour criterion discussed in Sect. \ref{fore} have been applied,
there are over $1000$ sources with $K> 18.0$~mag. This should be
sufficient to ensure that the Sobel filter can be reliably applied.

\begin{figure} 
\includegraphics[scale = 0.4]{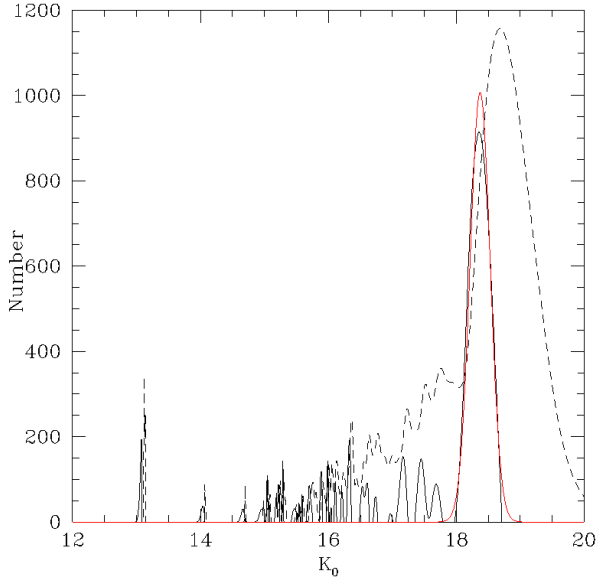}
\caption{\tiny{The smoothed $K$-band magnitude distribution (dashed line) 
of the sources in the $\sim 10' \times 10'$ central grid region. 
The Sobel filter (solid black line) has been applied and a Gaussian 
curve (solid red line) has been fitted to the strongest peak to 
determine the position of the TRGB magnitude and the associated 
error. The distribution is a generalised histogram and the vertical 
scale is arbitrary.}}
\label{sobel}
\end{figure}

Using the method described we locate the TRGB at $K = 18.37 \pm 0.15$~mag. 
However, \citet{2000A&A...359..601C} found the Sobel filter to be
systematically biased towards fainter magnitudes, due to the effects
of smoothing the data. Magnitude corrections were supplied by the same
authors and have been applied here, giving a revised TRGB magnitude of
$K = 18.28 \pm 0.15$~mag.  This is in reasonable agreement, within errors,
with previous measurements of the TRGB
\citep[e.g.][]{2011AJ....141..194G}. The derived TRGB value is also
consistent with the position of the discontinuity seen in the $K$-band
magnitude distribution presented in Fig. \ref{comp}.

We discuss
the variation of the TRGB magnitude across the face of IC~1613 in
Sect. \ref{results}, but for the purposes of isolating AGB sources in
our photometric sample we use a magnitude cut at $K = 18.28$~mag.

\subsection{Colour selection} 
\label{cols}

\subsubsection{C- and M-type AGB stars}
\label{colsel}
After isolating the AGB sources, it is necessary to classify them as
either C- or M-type before an estimate of the C/M ratio can be
made. C- and M-type stars can be identified by their position in the
($J-K$, $K$) CMD. M-type stars follow a vertical sequence above the
TRGB extending to brighter magnitudes over a relatively narrow range of
($J-K$) colours. On the other hand, C-type stars display a wider range
of colours which at ($J-K$)~$\sim 1$~mag overlap with, and at ($J-K$)~$\geq 1.2$~mag extend redder than, the M-type stars. In both Fig. \ref{CCDfg} and Fig. \ref{JKfg}, C-type stars result in an extended diagonal sequence
(depending on the scale of the axis). In Fig. \ref{JKfg} the separation between the two
sequences becomes more apparent at brighter magnitudes, however at
magnitudes $< 1$~mag brighter than the TRGB there is significant
overlap between the two spectral types, as can be seen in
Fig. \ref{JKfg}.

\begin{figure} 
\includegraphics[scale = 0.40]{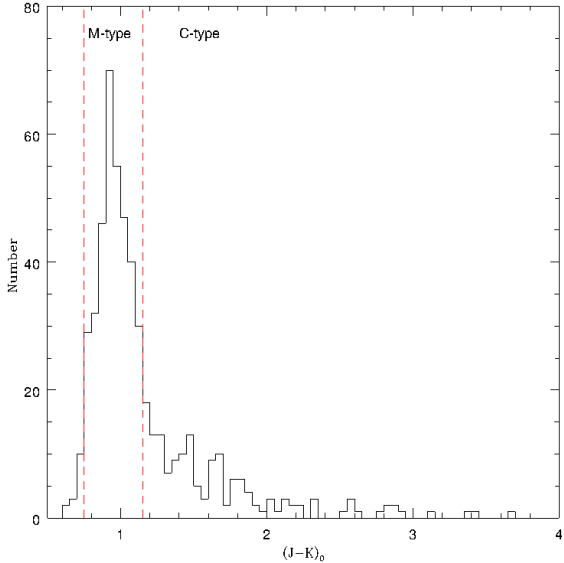}
\caption{\tiny{($J-K$) colour histogram of those sources with ($J-H$)~$\geq 0.64$~mag 
and brighter than the TRGB magnitude in the central grid region (Fig. \ref{grid}). The 
vertical dashed lines mark the position of the colour separation imposed here to 
distinguish between C- and M-type sources at ($J-K$)~$= 1.15 \pm 0.05$~mag and the 
blue limit at ($J-K$)~$= 0.75$~mag (see Sect. \ref{blue}).}}
\label{CMhist}
\end{figure}

From the CMD in Fig. \ref{JKfg}, the colour
separation between the C- and M-type stars can be estimated to lie
between ($J-K$)~$= 1.10-1.20$~mag. This estimate is refined using a
($J-K$) colour histogram of the AGB sources in the central grid region
(Fig. \ref{CMhist}). Due to the concentration of the M-type sources
over a narrow range of colours in comparison with the wider colour
distribution of the C-type population, the separation between the two
spectral types is marked by a discontinuity in the ($J-K$)
histogram. The strong peak in Fig. \ref{CMhist} identifies the M-type
population followed by a sudden decline in the number of sources and a
tail extending to redder ($J-K$) colours due to the C-type
population. The colour separation was found to lie at ($J-K$)~$=
1.15 \pm 0.05$~mag. For the purposes of comparison with other studies,
this translates into a colour separation of $(J-K_{s})_{2MASS} \sim
1.22$~mag on the 2MASS system using the transformations of
\citet{2001AJ....121.2851C}. 

This value has been used for the
purposes of classifying candidate C- and M-type AGB stars in our
photometric sample. However, as was noted by
\citet{2009A&A...493.1075B} and \citet{2013..InPreparation...S}, there
is no strict colour boundary between these two spectral types in
($J-K$) and the misclassification of AGB sources in both directions is
probable. Their broad range of colours means that C-type stars in
particular are likely to be seen on either side of this colour
boundary and the cut at ($J-K$)~$= 1.15$~mag may well represent the
red limit for the majority of the M-type population rather better than the blue limit
for the C-type population
\citep{2012A&A...537A.108K,2013..InPreparation...S}. Although in 
rare cases very M-type stars with ($J-K$) colours much redder than this 
have been found \citep{2011MNRAS.411.1597W}, such a colour cut
is appropriate for an initial estimate of the global C/M
ratio, as is the aim of this work. Spectroscopic data would be
required in order to refine the classification of individual
stars.

In their study of IC~1613, \citet{2009A&A...493.1075B}
conclude that a cut at ($J-K_{s}$)~$=1.40$~mag would provide a
reliable sample of C-type stars. From Fig. \ref{CMhist} we agree that
a cut around ($J-K$)~$\sim 1.30$~mag would provide the purest sample
of C-type stars, but it would also exclude a large number of bluer
C-type stars and so would not be appropriate for our purposes.

\subsubsection{Blue limit to M-type giants}
\label{blue}
After the application of the ($J-H$) foreground cut, those sources
brighter than the TRGB span a ($J-K$) colour range of $0.47 - 4.18$~mag (Fig. \ref{CMhist}). It is unlikely that the M-type AGB sources
would extend to such blue colours and it was therefore decided to
apply a `blue limit' to exclude contamination from K-type giants
\citep{1988PASP..100.1134B,2007A&A...474...35B}. From
Fig. \ref{CMhist} this limit is placed empirically at ($J-K$)~$= 0.75$~mag where we see a significant increase in the number of sources
contributing to the peak of what we believe to be M-type AGB
stars. Based on our $JHK$ photometric survey of NGC 6822
\citep{2012A&A...540A.135S} and the ($J-K$) colour distribution of
spectroscopically confirmed M-type giants in that galaxy
\citep{2012A&A...537A.108K}, this increase is expected to coincide
with the blue limit of the M-type giant ($J-K$) colour distribution.

The
final criteria used for the selection of AGB candidates in IC~1613
were ($J-H$)~$\geq 0.64$~mag and $K < 18.28$~mag. M-type sources were
then selected between $0.75 \leq$~($J-K$)~$< 1.15$~mag and C-type
sources were selected at ($J-K$)~$\geq 1.15$~mag. A red ($J-K$) limit
was not applied for the selection of the C-type candidates as these
sources have been shown to extend to least ($J-K$)~$\sim 2.4$~mag
\citep{2009A&A...493.1075B} in IC~1613, and the relatively few sources
in our C-type sample with colours redder than this may still be
genuine AGB stars that are heavily enshrouded by dust
\citep{2006MNRAS.370.1961Z}.

\section{Results}
\label{results}

\subsection{AGB catalogues}
\label{cats}
We present two catalogues: the first, Catalogue 1 (see Table
\ref{cat1}) contains the $843$ sources that we have classified as C-
or M-type AGB stars within $4.5$~kpc of the galactic centre. These
sources have been classified using the selection criteria derived in
Sect. \ref{anal} and are the basis for the C/M ratio and [Fe/H]
abundance derived below. In column $1$ we present an identification
number for the source followed by its position in Right Ascension (RA)
and Declination (Dec) for the equinox J2000 in cols. $2$ and
$3$. Cols. $4,5$ and $6$ contain the $J$-band magnitude, the
associated error and the quality flag classification in that band, the
same information is presented in cols. $7, 8, 9$ and $10, 11, 12$ for
the $H$- and $K$-bands respectively. Finally in Col. $13$ we list the source
classification, either C- or M-type. However, based on our findings in
Sect. \ref{fore} and Sect. \ref{stellden} its should be noted that
this catalogue is also likely to contain some foreground sources that
have been misclassified as giants. 

In Catalogue 2, we present all
those sources that met with our flag selection criteria for the full
observed area; no other selection criteria have been applied to this
catalogue. The first five lines of this catalogue are presented in
Table \ref{cat2}. With the exception of spectral type, the data in
each column is the same as for Table \ref{cat1}. The quality flag
classifications in columns $6,9$ and $12$ of both tables are as
follows; -9: saturated, -8: poor astrometric match, -3: compact
non-stellar, -2: probably-stellar, -1: stellar, 0: noise-like and 1:
non-stellar.

\begin{table*} 
\centering                       
\begin{tabular}{cccccccccccccc}
\hline
ID &$\mathrm{RA}$ &$\mathrm{Dec}$ &$J$ &$J \mathrm{-error}$ &$J \mathrm{-flag}$ &$H$
&$H \mathrm{-error}$ &$H \mathrm{-flag}$ &$K$ &$K \mathrm{-error}$ &$K \mathrm{-flag}$ &Sp. Type\\ 
   &$\mathrm{(deg)}$ &$\mathrm{(deg)}$ &$\mathrm{(mag)}$ &$\mathrm{(mag)}$ &
&$\mathrm{(mag)}$ &$\mathrm{(mag)}$ & &$\mathrm{(mag)}$
&$\mathrm{(mag)}$ & &\\     
\hline
1575 &16.151642 &1.865478 &16.39 &0.0070 &-1.0 &15.74 &0.01 &1.0  &15.61 &0.01 &-1.0 &M\\
1971 &16.256805 &1.8747   &19.86 &0.128  &-1.0 &19.02 &0.11 &-1.0 &18.21 &0.09 &-1.0 &C\\
2089 &16.250389 &1.880472 &19.26 &0.075  &-1.0 &18.56 &0.08 &-1.0 &17.74 &0.06 &-1.0 &C\\
2176 &16.205189 &1.888278 &19.88 &0.124  &-1.0 &18.88 &0.10 &-1.0 &18.14 &0.08 &-1.0 &C\\
2260 &16.162148 &1.893214 &18.31 &0.032  &-1.0 &17.63 &0.03 &-1.0 &17.47 &0.05 &1.0  &M\\
... &... &... &... &... &... &... &... &... &... &... &...&...\\
\hline\\
\end{tabular}
\caption[]{The first five lines of Catalogue 1, see Sect. \ref{cats} for more information
on the table contents.}   
\label{cat1} 
\end{table*}

\begin{table*} 
\centering                       
\begin{tabular}{ccccccccccccc}
\hline
ID &$\mathrm{RA}$ &$\mathrm{Dec}$ &$J$ &$J \mathrm{-error}$ &$J \mathrm{-flag}$ &$H$
&$H \mathrm{-error}$ &$H \mathrm{-flag}$ &$K$ &$K \mathrm{-error}$ &$K \mathrm{-flag}$\\ 
 &$\mathrm{(deg)}$ &$\mathrm{(deg)}$ &$\mathrm{(mag)}$ &$\mathrm{(mag)}$ &
&$\mathrm{(mag)}$ &$\mathrm{(mag)}$ & &$\mathrm{(mag)}$
&$\mathrm{(mag)}$ & \\     
\hline
12  &15.864609 &1.687656 &18.53 &0.04 &1.0  &18.19 &0.05 &-1.0 &17.90 &0.07 &-1.0\\  
34  &16.268135 &1.6881   &0.0   &0.0  &0.0  &18.50 &0.07 &-1.0 &18.21 &0.09 &-1.0\\ 
95  &15.965066 &1.689939 &18.30 &0.03 &1.0  &17.85 &0.04 &-1.0 &17.54 &0.05 &-1.0\\  
108 &16.052319 &1.69022  &18.02 &0.03 &1.0  &17.64 &0.03 &-1.0 &17.25 &0.04 &-1.0\\  
132 &16.110607 &1.690867 &19.55 &0.09 &-1.0 &19.13 &0.12 &-1.0 &18.56 &0.12 &1.0\\ 
... &... &... &... &... &... &... &... &... &... &... &...\\
\hline\\
\end{tabular}
\caption[]{The first five lines of Catalogue 2, see Sect. \ref{cats} for information
on the table contents.}   
\label{cat2} 
\end{table*}

\subsection{Spatial distributions}
\label{spacdist} 

\subsubsection{The removed foreground and K-type giants}
\label{removeFG}
\begin{figure*} 
\includegraphics[scale = 0.61]{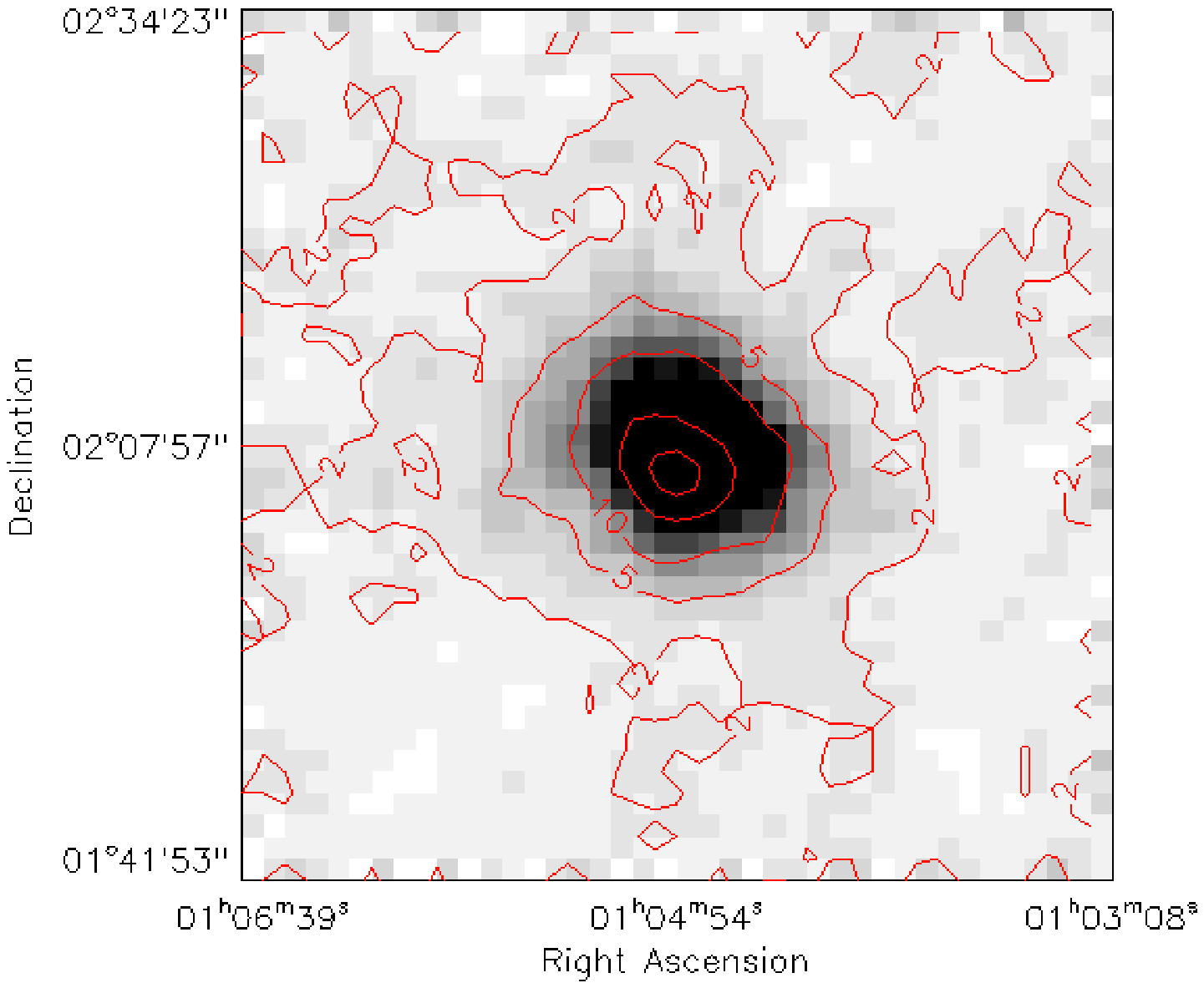}
\includegraphics[scale = 0.38]{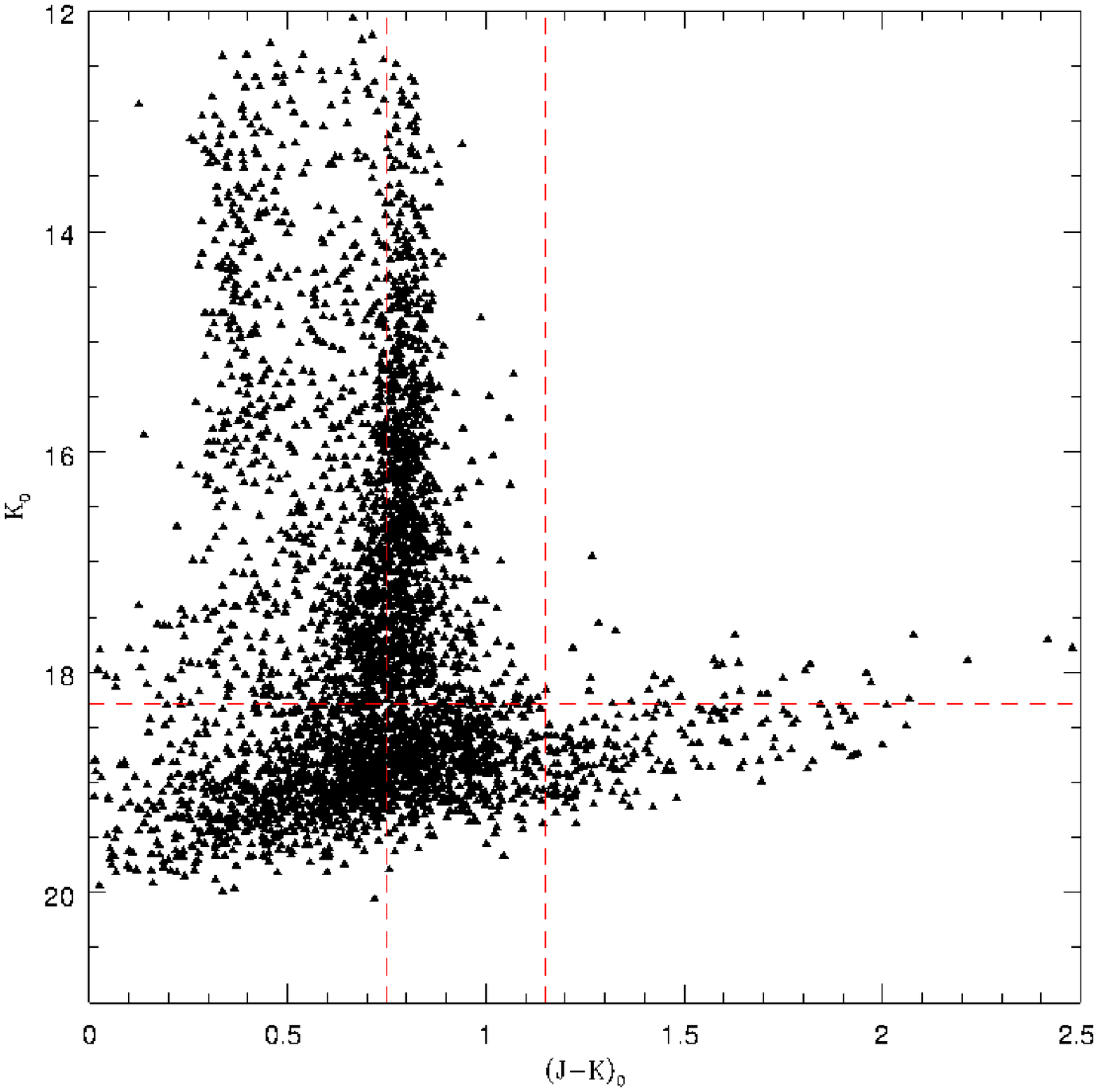}
\caption{\tiny{Left: Density distribution of the foreground and IC~1613 K-type giant sources removed from the $JHK$ data set as described in Sect. \ref{fore}. Using $1600$ bins across the full observed area contours are at $2$, $5$, $10$, $25$ and $40$ per arcmin$^{2}$. Right: CMD of all the sources removed as foreground and the K-type giants. The dashed horizontal and vertical lines are the AGB selection criteria discussed in Sect. \ref{anal}.}} 
\label{FGspace}
\end{figure*}

In the left panel of Fig.\ref{FGspace} we present a surface density
plot of those sources that have been removed using the
($J-H$) colour criterion. The overdensity in the centre indicates that
many genuine IC~1613 sources have been removed by the colour cut. This
was expected, as given the overlap in colour between K-type giant and
foreground sources \citep{2000ApJ...542..804N}, foreground removal by
colour selection will always be imperfect. The question we must
address is what impact this will have on the estimated numbers of C-
and M-type giants in IC~1613.

In the right panel of
Fig.\ref{FGspace} we show a CMD of the sources removed from our data
set as foreground. Nearly half of the sources removed lie below
the TRGB and so would not contribute to the estimated number of AGB
stars, genuine IC~1613 sources in this region of the CMD are probably RGB
stars. The majority of the removed sources above the TRGB form a
vertical sequence around the blue limit at ($J-K$)~$= 0.75$~mag and are probably K-type giants rather than M-type which tend to have redder
colours (see Fig. \ref{JKfg}). Based on this CMD, the impact of our
foreground cut on the number of C-type sources will be negligible, and
while the overlap in colours means that it is inevitable that some M-type giants will be removed with the foreground, the majority of
the genuine IC 1613 giant sources removed are thought to be K-type and RGB stars rather than M-type giants. The removal these sources is of no consequence to our work; therefore, although the C/M
ratio we derive will be affected by the removal of some genuine
M-type giant sources, as well as the inclusion of a small number of IC~1613 K-type giants and foreground contaminants, we do not expect
the impact to be severe. However, we do consider the effects of alternative cuts in ($J-K$) and ($J-H$) in Sect. \ref{senJK} and Sect. \ref{senJH}.

\subsubsection{Stellar density profiles}
\label{stellden}
\begin{figure} 
\includegraphics[scale = 0.40]{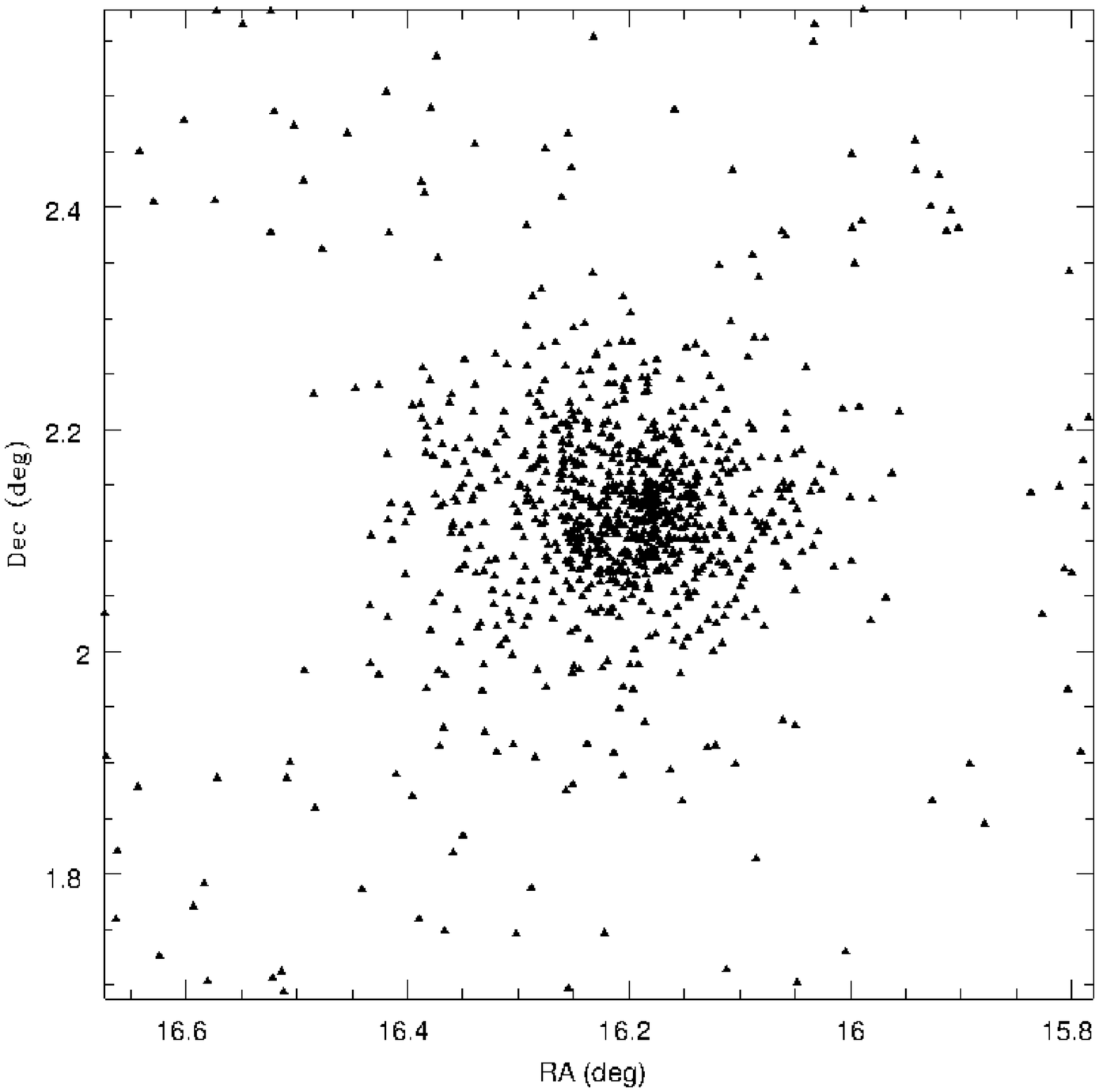}
\includegraphics[scale = 0.40]{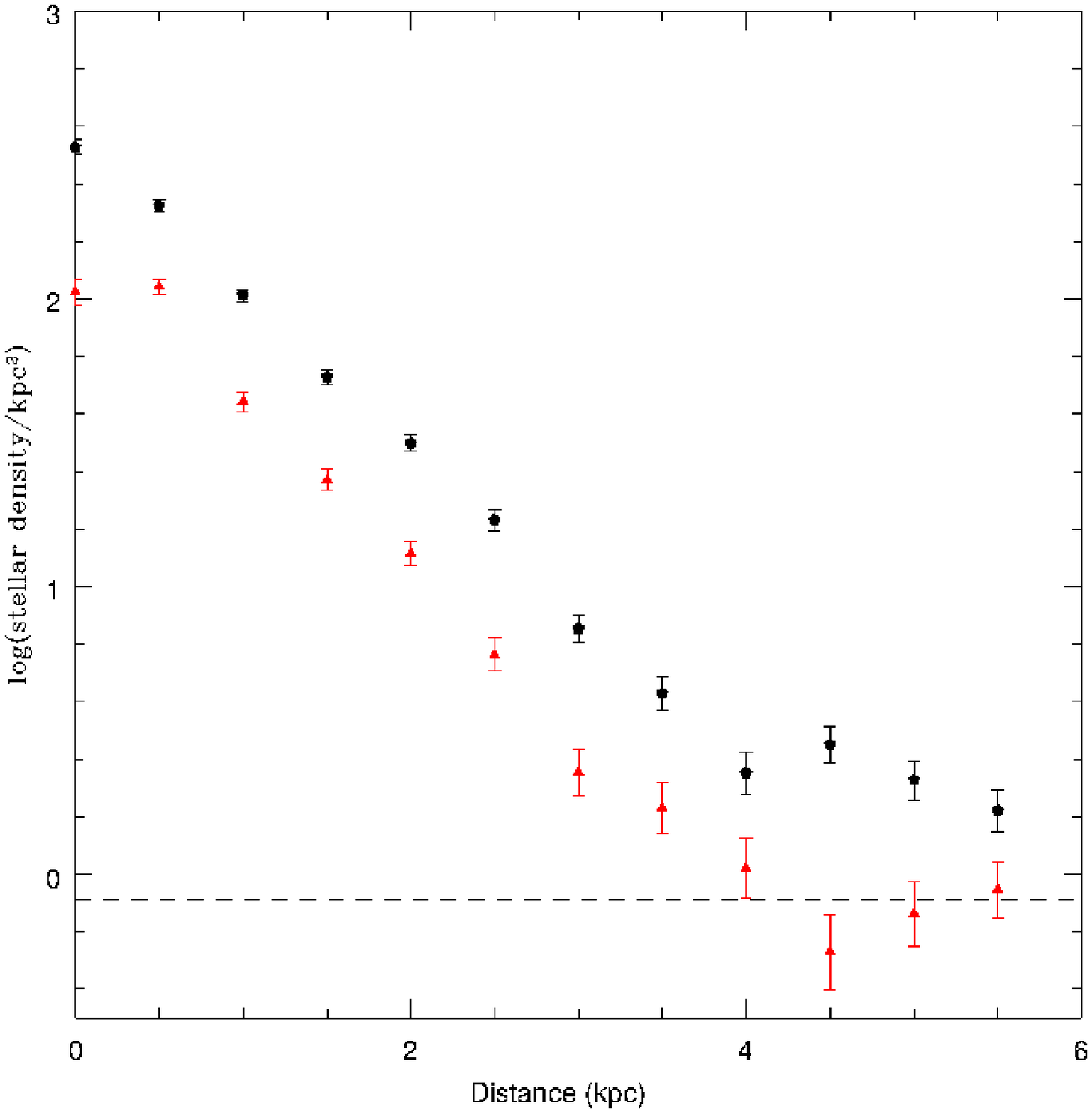}
\caption{\tiny{Top: Spatial distribution of the candidate 
AGB stars in the observed area. Bottom: Log plot of 
the number of AGB (red triangles) and RGB (black 
dots) classified sources per~kpc$^{2}$ as a function of
distance from the galactic centre. The dashed line at $-0.086$ 
indicates the residual background level brighter 
than the TRGB, estimated from the four corners of 
the grid (see Sect. \ref{trgb})}} 
\label{seanstell}
\end{figure}

In the top panel of Fig.\ref{seanstell} we show the spatial
distribution of the AGB candidates we have identified. There is a clear overdensity in
the centre, with the number of sources declining with distance from
the galactic centre. Sources in the outer part of the galaxy may be
genuine AGB sources or they may be residual foreground
contamination within our sample, estimated in Sect. \ref{fore} as
$\sim 0.8$~stars per~kpc$^{2}$. In order to examine
the distribution of our candidate sources in more detail and to make a
second estimate of the level of remaining foreground contamination in our sample, we
plot the density of candidate sources per kpc$^{2}$ as a function of distance from the
galactic centre.

As IC~1613 is inclined to the line of sight, we
first de-projecting our sources on to a flat plane using a position
angle (PA) of $58^{\circ}$ and an inclination of \textit{i}~$=
38^{\circ}$, as calculated by \citet{1989AJ.....98.1274L} based on the
H~I distribution of the galaxy. The distance in~kpc from the galactic
centre in the plane of the galactic disk is then calculated for every
source. The number of RGB and AGB sources is then calculated in $12$
annuli at intervals of $0.5$~kpc between $0-6$~kpc from the centre of
the galaxy. The results are presented in the bottom panel of
Fig. \ref{seanstell}. We limit our outer-most annulus to $6$~kpc 
($\sim 27'$), as based on estimates of the tidal radius of IC~1613 by 
\citet{1991ApJ...369..372H} and \citet{2007A&A...466..875B} no significant 
number of stellar sources are expected beyond this.

In
Fig. \ref{seanstell} we see a steep decline in the number of AGB
sources out to $\sim 4 - 4.5$~kpc before the profile levels off
between $\sim 4 - 6$~kpc. The RGB density profile shows similar
behaviour. A distance of $4 - 4.5$~kpc from the galactic centre
translates into an angular distance of $18 - 20'$ at a distance of
$758$~kpc. These measurements are consistent with previous estimates
by \citet{2007AJ....134.1124B} who report the detection of RGB stars
out to $r > 16.5'$ and \citet{2007A&A...466..875B} who found that the
density profile of giant stars, mainly on the RGB, declined to zero at
$r \sim 23'$. \citet{2007A&A...466..875B} also confirmed the findings
of \citet{2002A&A...394...33T} that no giants belonging to IC~1613 are
seen in the HST archive images at $27'$ and $33'$, $\sim 6$ and $\sim
7$~kpc respectively. This suggests that the AGB profile in
Fig. \ref{seanstell} is declining as it approaches the edge of the
galaxy and that beyond $\sim 4-4.5$~kpc the candidate AGB sources we
detect have a high probability of being foreground interlopers. The
level of AGB contamination inferred from Fig. \ref{seanstell} is
consistent with that inferred from the grid corners in
Sect. \ref{trgb}.

\begin{figure} 
\includegraphics[scale = 0.4]{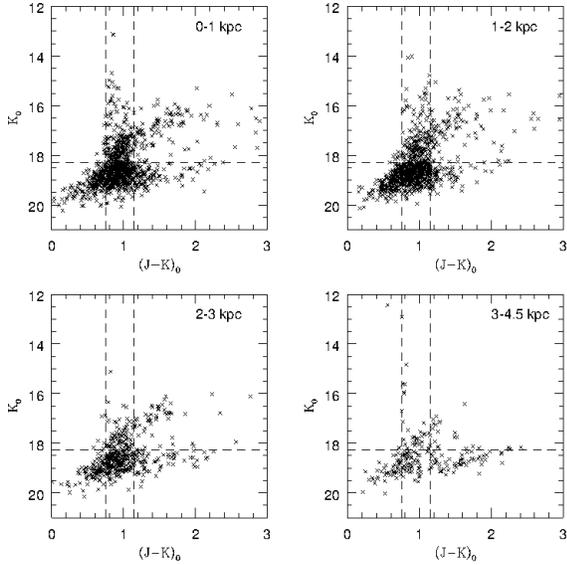}
\caption[]{\tiny{CMDs of those sources with 
($J-H$)~$\geq 0.64$~mag within each of the 
annuli between $0-1$~kpc, $1-2$~kpc, $2-3$~kpc and $3-4.5$~kpc. Foreground stars with ($J-H$)~$\geq 
0.64$~mag will also be present. The dashed 
horizontal and vertical lines mark the 
position of the TRGB and our ($J-K$) 
classification criteria respectively.}}
\label{CMDgrid}
\end{figure}

In Fig. \ref{CMDgrid} we present four CMDs of the sources between $0-1$~kpc, $1-2$~kpc, $2-3$~kpc and $3-4.5$~kpc. Beyond $3$~kpc the
vertical sequence defined by the M-type stars is no longer visible and
the C-type branch is much weaker. This supports our earlier
conclusions about the decreasing stellar density approaching $4.5$~kpc. We
therefore limit our selection of AGB stars in IC~1613 to within
$4.5$~kpc of the galactic centre. The number of candidate AGB C- and
M-type stars per~kpc$^{2}$ beyond this limit has then been used to
make an estimate of the number of remaining foreground contaminants in
our sample before making a final estimate of the C/M ratio in
Sect. \ref{c/m}.

\subsection{The TRGB}
\label{trgb2}
The TRGB has been shown to be sensitive to both age and metallicity,
becoming brighter in older sources within a population of a single
metallicity but moving to fainter magnitudes with decreasing
metallicity in a population of a single age. In a galaxy such as
IC~1613 which has undergone continuous star formation the opposing
effects of these population characteristics can make it difficult to
interpret any change in the measured TRGB magnitude
\citep{2005MNRAS.357..669S}, however in conjunction with the derived
[Fe/H] abundance below, any variation in the TRGB may help us to
better understand the age and metallicity distribution of the giant
population in IC~1613.

\begin{figure} 
\includegraphics[scale = 0.40]{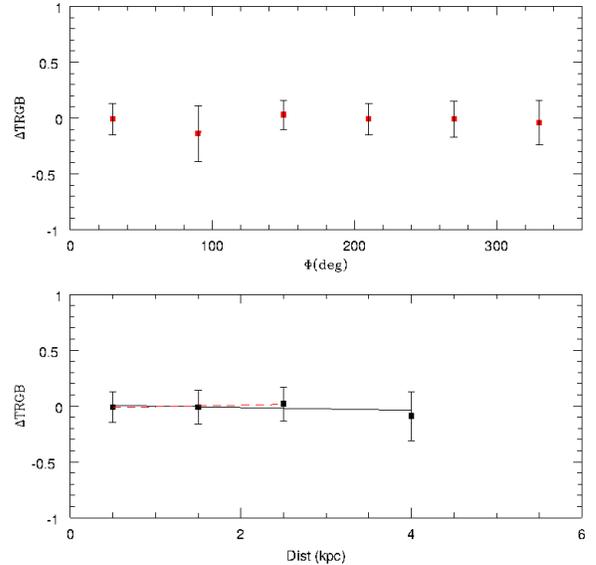}
\caption[]{\tiny{Variation of TRGB magnitude as 
a function of azimuthal angle (top) and as a function 
of distance from the galactic centre (bottom). In 
the top panel, the TRGB has been measured only in the 
$6$ of the inner annuli seen in Fig. \ref{circ}. In 
the bottom panel the solid black line shows a weighted 
linear fit all all the data points, the dashed red line 
shows a weighted fit to only the first $3$ data points.}}
\label{trgb3}
\end{figure}

In Sect. \ref{anal} the TRGB was found to be
$K = 18.28 \pm 0.15$~mag. The area within the $4.5$~kpc limit was
subdivided into $4$ annuli between $0-1$~kpc, $1-2$~kpc, $2-3$~kpc and
$3-4.5$~kpc to establish if there is any variation in the TRGB with
increasing distance from the galactic centre. The final annuli is has
a larger radius to ensure that the criteria for the reliable
application of the Sobel filter were met (Sect. \ref{trgb}). The
results of these measurements are shown in the bottom panel of
Fig. \ref{trgb3} which gives $\Delta$TRGB~$= K_{TRGB} - 18.28$. As can been seen,
there is very little variation in the measured TRGB with distance from
the galactic centre. Overall there is a spread of $\Delta K = 0.11$~mag, however if the last data point, which had only $136$ sources in
the magnitude range $K > 18.0$~mag, is excluded this spread is reduced
to $\Delta K = 0.03$~mag. An error weighted fit to all of the data points
shows a shallow negative slope ($\Delta$TRGB $ = -0.01 (\pm 0.02)
\times d + 0.01 (\pm 0.04)$), where $d$ is the galactocentric distance
in~kpc. This suggests that the TRGB magnitude increases with distance
from the galactic centre but at the level of only $0.5 \sigma$,
i.e. such that we can not conclude that the TRGB is varying with
distance.

\begin{figure} 
\includegraphics[scale = 0.40]{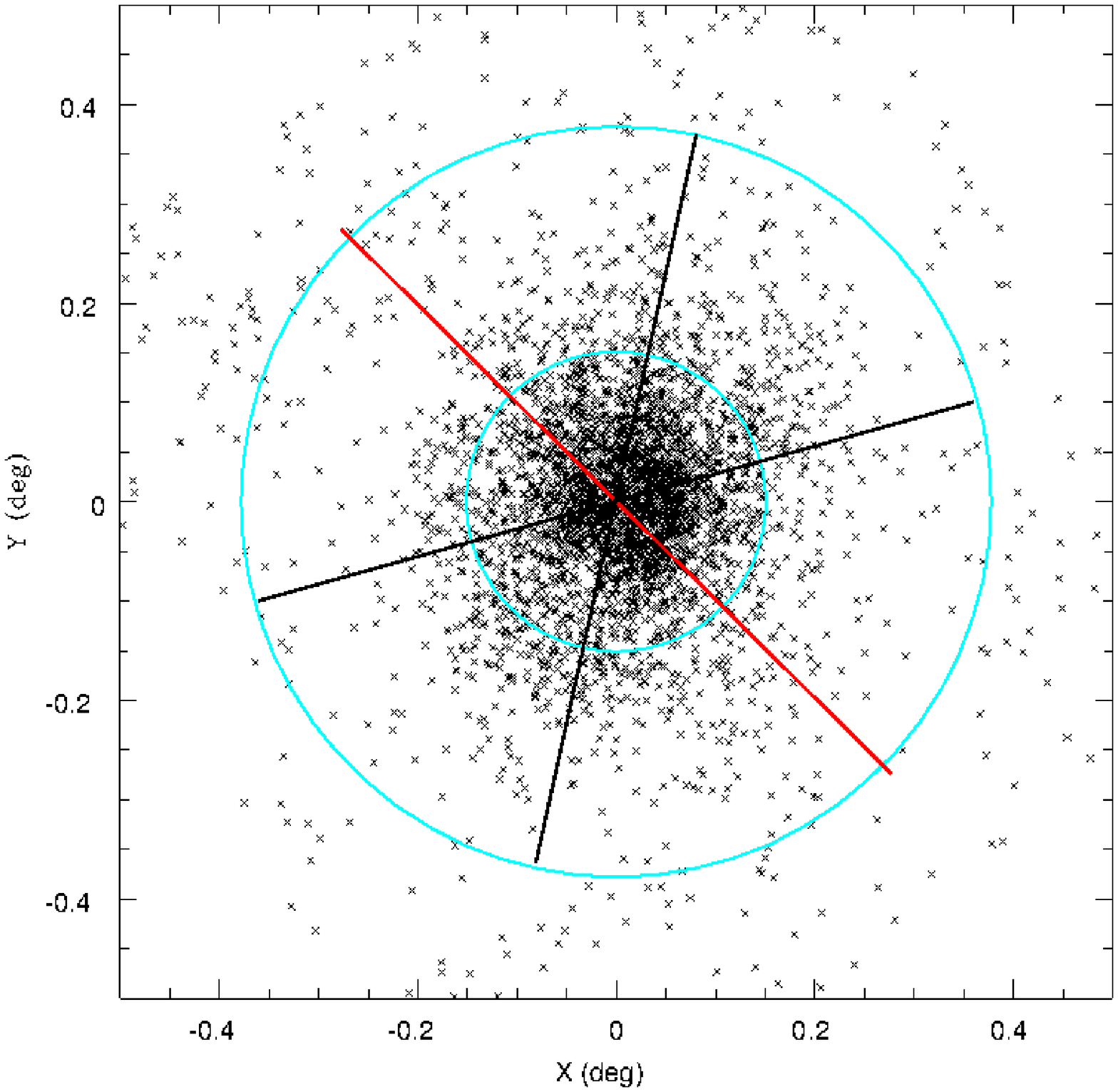}
\caption[]{\tiny{The observed area has been 
aligned with the PA ($58^{\circ}$) of the galactic 
H~I and de-projected using an inclination of 
$38^{\circ}$. The black crosses show the positions 
of those sources with ($J-H$)~$\geq 0.64$~mag. 
Two circles at radius $2$~kpc 
($9'$) and $4.5$~kpc ($20'$) have been imposed 
and the area divided into $6$ segments in each annuli  
order to examine the TRGB, C/M ratio and [Fe/H] 
abundance within IC~1613 as a function of angle. 
The thick red line marks the orientation of the 
approximate position of the PA in the de-projected 
image.}}
\label{circ}
\end{figure}

In addition to the $4$ annuli described above, the area within
$4.5$~kpc was separately divided into $12$ segments, $6$ between
$0-2$~kpc and $6$ between $2-4.5$~kpc each with a separation of
$60^{\circ}$ (see Fig. \ref{circ}), in order to study the TRGB as a
function of azimuthal angle. The measured TRGB magnitudes in the $6$
inner segments are presented in the top panel of Fig. \ref{trgb3}. The
outer segments unfortunately did not have enough sources with magnitude $K > 18.0$~mag for the reliable application of
the Sobel filter. Overall there was a spread of $\Delta K = 0.17$~mag
in the measured TRGB in the $6$ segments but as can be seen in
Fig. \ref{trgb3} there is no overall trend in the TRGB and the average
value of $K = 18.25 \pm 0.07$~mag is in excellent agreement with the
TRGB magnitude we use for the selection of AGB sources.

\subsection{The C/M ratio}
\label{c/m}
Within $4.5$~kpc of the galactic centre of IC~1613 (in the de-projected plane), from $291$
C-type and $552$ M-type AGB stars we infer C/M~$ = 0.53 \pm 0.04$. This value
is calculated without any adjustment for the remaining foreground
contamination, which from the number of C- and M-type classified stars
we find beyond $4.5$~kpc we calculate to be $\sim 0.28$ C-type and
$\sim 0.41$ M-type sources per~kpc$^{2}$. However, the low level of
foreground contamination in the direction IC~1613 means that if a statistical adjustment is made to account for the remaining
contaminants, it has very little impact on the derived C/M ratio which
becomes $ 0.52 \pm 0.04$. Using the C/M vs. [Fe/H] relation of \citet{2009A&A...506.1137C}
([Fe/H]~$= -1.39 \pm 0.06 - (0.47 \pm 0.10)$log$(C/M)$) we derive a
global [Fe/H] abundance of $ -1.26 \pm 0.07$~dex. 

\begin{figure*} 
\includegraphics[scale = 0.45]{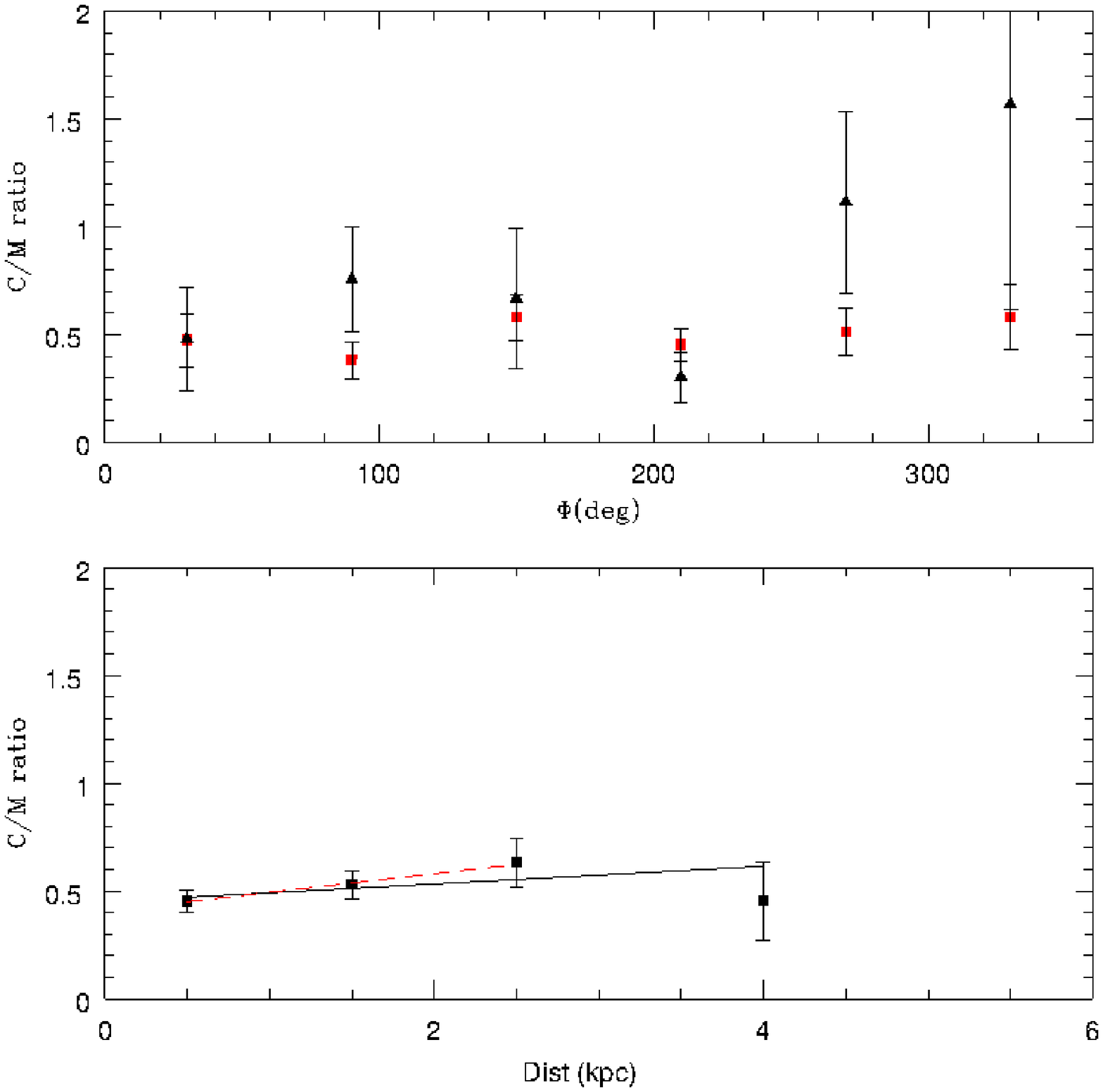}
\includegraphics[scale = 0.45]{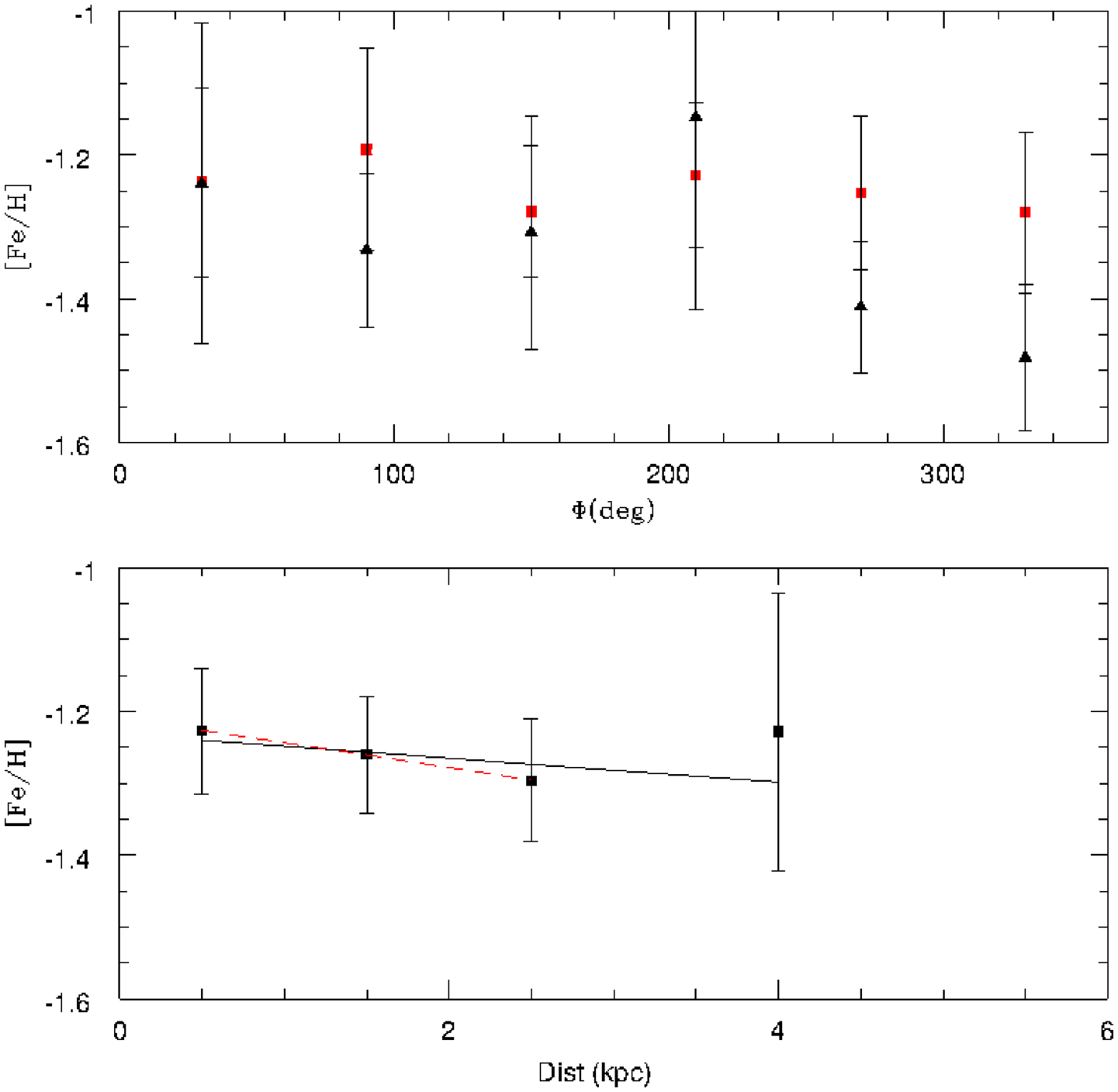}
\caption{\tiny{Left: In the top panel the 
C/M ratio is plotted as a function of azimuthal 
angle for each of the segments shown in Fig. 
\ref{circ}. The inner segments, $0-2$~kpc (red 
squares), and the outer segments $2-4.5$~kpc (black 
triangles) are shown separately. The bottom 
panel shows the C/M ratio as a function of distance 
from the galactic centre. The solid black line 
is a weighted linear fit to all four points, the 
red dashed line is also weighted but is only 
fitted to the first three points. Right: The 
same information is presented in the top and 
bottom of the right hand panel for the [Fe/H] 
abundance, derived from the C/M ratio using the 
relation of \citet{2009A&A...506.1137C}.}} 
\label{Dists}
\end{figure*}

The same $4$ annuli between $0 - 4.5$~kpc and the same $12$ segments shown in
Fig. \ref{circ} that were used in Sect. \ref{trgb2}, have also been used to study the C/M ratio and the [Fe/H]
abundance as a function of radial distance and azimuthal angle. Our
results are presented in Fig. \ref{Dists}. In the bottom left panel of
Fig. \ref{Dists} we see a spread of $0.18$ in the C/M ratio out to
a radius of $4.5$~kpc. A weighted fit to the data ($\Delta$(C/M)~$= 0.04
(\pm 0.04) \times d + 0.45 (\pm 0.05)$) shows a shallow trend towards
a greater value of C/M (i.e. lower metallicity) with increasing
distance from the galactic centre but of only $1 \sigma$
significance. The spread in C/M translates into [Fe/H] between $-1.23$
and $-1.30$~dex. However, a weighted fit to all of the data points
($\Delta$[Fe/H]~$= -0.02 (\pm 0.02) \times d -1.23 (\pm 0.03)$) does
not suggest a strong trend to lower metallicities with increasing
galactocentric distance.

The variation in the calculated C/M ratio and the [Fe/H] abundance as
a function of azimuthal angle for each segment in Fig. \ref{circ} are
shown in the top panels of Fig. \ref{Dists}. The angle is measured
anti-clockwise from the PA ($58^{\circ}$) of the galaxy. The
measurements for the inner segments between $0-2$~kpc are shown by
the red squares and those for the outer segments between $2-4.5$~kpc
are shown by the black triangles. A spread of $\Delta$(C/M)~$= 0.20$
is seen in the measurements from the inner segments while a much
larger spread of $\Delta$(C/M)~$= 1.27$ is seen in the measurements
from the outer segments, however the statistical errors on the outer segments are
large. We therefore restrict our analysis to the more reliable inner
segment measurements. The measured C/M seems to show a slight
sinusoidal variation with angle around the galaxy. The variation in
C/M in the inner segments translates into a spread of $\Delta$[Fe/H]~$= 0.09$~dex, and as shown in the top right panel of Fig. \ref{Dists}, the
derived [Fe/H] values reflect the sinusoidal variation of the C/M
ratio although any fit to the data would have a very small amplitude
and would not represent a significant variation in metallicity as a
function of angle.

\section{Discussion}
\label{finish}

\subsection{The C/M ratio: age-metallicity dependence}
\label{agemet}
The analysis in Sect. \ref{anal} has been conducted with the aim of isolating and classifying the AGB population of IC~1613 in order to derive the global metallicity of the galaxy using the C/M ratio, as we have done in Sect. \ref{results}. This reflects a traditional interpretation of the relationship between the C/M ratio and the metallicity of the interstellar medium, that while sufficient for our purpose of providing an overview of the galactic metallicity distribution, is ultimately too simplistic. In \citet[Sect. 4.1.1]{2012A&A...540A.135S} we provided a detailed discussion of the importance of population age when interpreting the C/M ratio and some important results in the literature at the time of publication. We briefly summarise that discussion here and provide an overview of other recent results that aim to further constrain our understanding of the TP-AGB phase.

As we outlined in Sect. \ref{intro} the creation of intrinsic C-type stars during the TP-AGB phase results from the dredging-up of material, primarily ($^{12}$C), from the inter-shell region to the surface. This process, known as the third dredge-up (TDU), subsequent changes in the molecular opacity of the stellar atmosphere and the resulting change in the rate of mass loss, as well as other processes such as hot bottom burning (HBB) are believed to severely limit the mass, and therefore age, range over which C-type AGB stars can form \citep{2003MNRAS.338..572M,2010ApJ...724.1030G,2010MNRAS.tmp..431H}. Several observational and theoretical studies have attempted to place constraints on this mass (age) range for different populations. Using detailed stellar models with a range of metallicities (Z~$= 0.004, 0.008, 0.02$) \cite{2003PhDT..01..286K} concluded that only AGB stars more massive than $1-1.5$~M$_{\odot}$ will undergo TDU and hence could become C-type stars. \citet{2010ApJ...724.1030G} obtained similar results for even more metal poor stars (Z~$= 0.001$) using their own models and calibrating the TP-AGB tracks against HST observations of AGB stars in nearby galaxies. \citet{2014ApJ...782...17K} also used theoretical models calibrated on observational data to examine the evolution of TP-AGB stars at solar metallicity and concluded that the TDU will only occur in stars with an initial mass of $\geq 2$~M$_{\odot}$. Importantly, they also concluded that at this metallicity the fraction of the TP-AGB lifetime spent as a C-type star is at a maximum for stars with an initial mass of $2.6$~M$_{\odot}$ but still only accounts of $\sim 23\%$ of the time spent in that phase of evolution. \citet{2012MNRAS.424.2345V} were able to put an upper limit on the mass range over which C-type stars would form; following the evolution of stars with initial mass $1 - 8$~M$_{\odot}$ and metallicity Z~$=0.008$ through the TP-AGB phase, they concluded that only stars with an initial mass of $\leq 3.5$~M$_{\odot}$ would become C-rich. This result is consistent with the earlier findings of \citet{2010MNRAS.408.2476V}, who reached the same conclusion for stars with a metallicity of Z~$= 0.001$. However, \citet{2014MNRAS.438.1741F} found that depending on the treatment of molecular opacity with successive dredge-up events stars with an initial mass of $5-6$~M$_{\odot}$ and metallicity Z~$= 0.001$ can achieve an atmospheric C/O ratio of greater than unity. While even stars with significantly higher metallicity (Z~$= 0.02$) and initial mass $5$~M$_{\odot}$ can attain a surface C/O ratio that approaches, although never achieves, unity. \citet{2013ApJ...774...83B} were able to put a metallicity ceiling on the formation of C-type stars. During a trial of a new tool for the detection of C- and M-type, \citet{2013ApJ...774...83B} found only one C-type star in a small field near the centre of M31, a result which they attributed to the relatively high metallicity ([M/H]~$\sim 0.1$~dex) of that region.

The dependence of the C/M ratio on the age, as well as the metallicity, of the population and the anti-correlation between these two characteristics can significantly complicate the interpretation of the measured C/M ratio. For example, following conventional wisdom a `young' AGB population (e.g. $\leq 4$~Gyr) is expected more metal rich than an older AGB population (e.g. $\geq 6$~Gyr) in the same galaxy (assuming efficient interstellar mixing), and this would be reflected in a lower C/M ratio for that population. However, given the limited mass range over which C-type stars are expected to evolve, a higher C/M ratio may in fact be observed in the younger population than in the older population which, although more metal poor, no longer contains any stars of sufficient mass to become C-type stars. \textbf{The application of the single variable C/M~vs.~[Fe/H] relations currently available in the literature would}, in this instance, lead to the erroneous conclusion that the younger population was more metal poor \citep[e.g.][]{2010MNRAS.tmpL.129F}. Useful plots have been presented by \citet{2008MNRAS.388.1185G} and \citet{2010MNRAS.tmp..431H} clearly demonstrating how the number of C-type stars in a population can change with time, and \citet{2006A&A...448...77C} present an instructive plot showing the dependence of the C/M ratio on the age of the underlying population.

The mechanisms we listed above and their dependence on mass and metallicity produce the same dependencies in the C/M ratio. Unfortunately, they are also among the most poorly understood mechanisms of the AGB phase. At present there is no calibration of the C/M~vs.~[Fe/H] relation, that we are aware of, that takes the age dependence of the C/M ratio into account. We do not have sufficient data to explore the age and mass distribution of our candidate sources. We have therefore relied on the classical interpretation of the C/M ratio in this work and have brought the age dependence of the C/M ratio to the readers notice to aid in any future reinterpretation of our results. However, we note that the C/M values we obtain with increasing galactic radius are very consistent, showing only a $1 \sigma$ variation. If a significant age or metallicity gradient were were present with increasing galactic radius we would expect to see a more significant change in the C/M ratio. Otherwise a fine balance between the age and metallicity components of the population would be required to cancel the effect on the C/M ratio.

\subsection{\citet{2000AJ....119.2780A}}
\label{Albert}
Several authors have made use of optical photometry to study the intermediate age stars in IC~1613 \citep[e.g][]{1988AJ.....96.1248F,1999AJ....118.1657C,2002A&A...394...33T,2007A&A...466..875B}. One such study by \citet{2000AJ....119.2780A} made use of a multi-filter technique known as the $CN-TiO$ method \citep{1982AJ.....87.1739P, 1986ApJ...305..634C} to detect and classify AGB stars in particular. The $CN - TiO$ method relies on two broad-band optical filters and two narrow-band (CN and TiO) filters to classify individual sources and has been shown to be highly successful in the classification of C-type stars in particular \citep{1996AJ....112..491B,2013..InPreparation...S}. Using this method \citet{2000AJ....119.2780A} identified $195$ candidate C-type AGB sources in IC~1613, within the $4.5$~kpc limit we have imposed and estimated the global C/M ratio in the galaxy to be $0.64$.

However, although optical studies have successfully detected AGB stars it has been noted by \citet{1995AJ....109.2480B} and \citet{2009A&A...504.1031G} that as the environment around an AGB star becomes increasing polluted by dust, the shift in its spectral energy distribution means it can go undetected at optical wavelengths. This can be a problem for C-type stars in particular leading to higher levels of incompleteness for these stars. The problem of detecting AGB stars at optical wavelengths in IC~1613 was clearly demonstrated by \citet{2007ApJ...667..891J}, in one of a series of papers that used images taken by the Spitzer Infrared Camera Array (IRAC) to conduct a census of AGB stars in LG dIrr galaxies. When comparing their IR data to the optical data of \citet{2000AJ....119.2780A}, \citet{2007ApJ...667..891J} found that \citet{2000AJ....119.2780A} had detected only $50\%$ of the AGB population. Furthermore, \citet{2007ApJ...667..891J} estimated that as \citet{2000AJ....119.2780A} applied a blue limit in ($R-I$) when selecting AGB candidates they actually used only $18\%$ of the AGB population when calculating the global C/M ratio. The poor AGB detection rate reported for \citet{2000AJ....119.2780A} was repeated when comparing the IRAC data to other optical data sets in IC~1613 and in other dwarf galaxies \citep{2007ApJ...656..818J,2009ApJ...697.1993B}. As a result of the difficult of detecting redder AGB stars at optical wavelengths recent studies \citep[e.g][]{2000RMxAA..36..151B, 2009JASS...26..421J} have made use of NIR photometry to detect AGB stars; as these stars emit most of their light at these wavelengths \citep{1990ApJ...352...96F} they brightest are among the brightest objects in the NIR.

\textbf{The NIR data we present here are among the deepest available for IC~1613 and, based on the results of \citet{2007A&A...466..875B}, the field of view covers the whole galaxy. However, as we have demonstrated in Sect. \ref{anal} \& \ref{results}, the selection and classification of AGB sources at NIR wavelengths is not simple and is prone to error at colour and magnitude boundaries}. It was decided to cross-match our AGB candidate sources with those of 
\citet{2000AJ....119.2780A} in order to study the $JHK$ colours and magnitudes 
of an incomplete but highly reliable C-type star sample and to estimate the potential level of 
error in our derived C/M ratio.

\begin{figure*} 
\includegraphics[scale = 0.45]{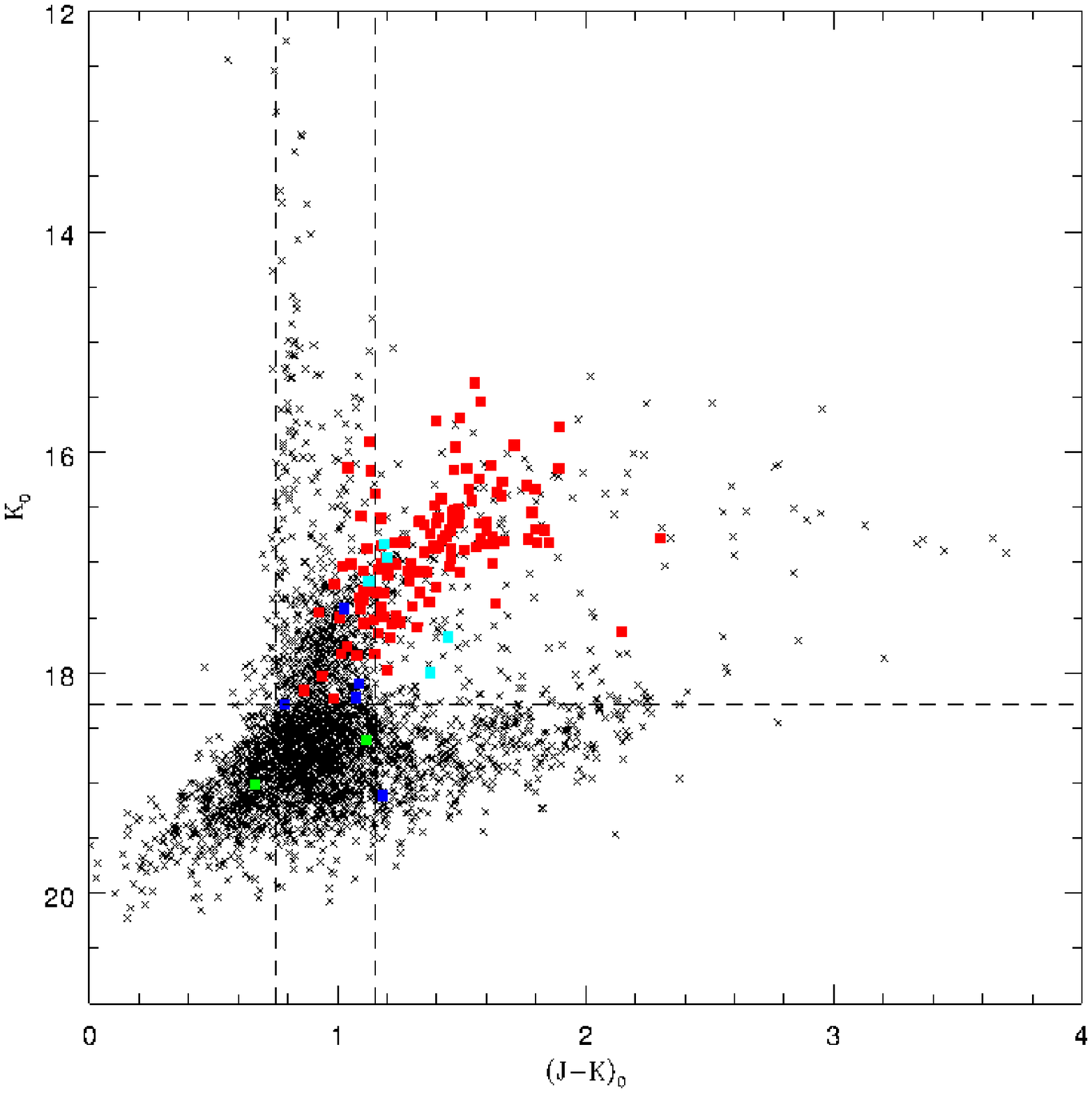}
\includegraphics[scale = 0.45]{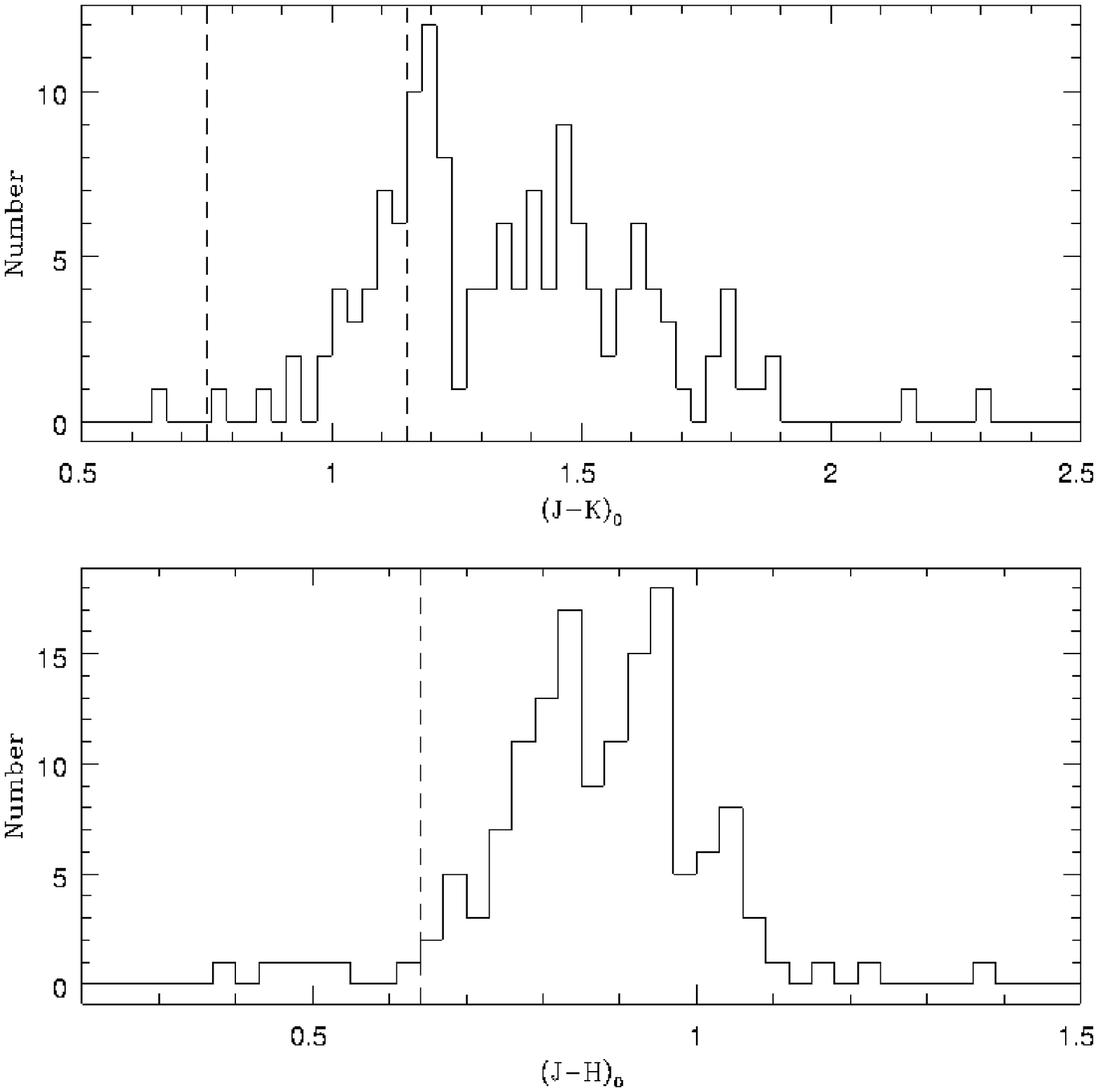}
\caption{\tiny{Left: CMD of all the sources (black crosses) in the original photometric data that met the quality flag criterion and have ($J-H$)~$>0.64$~mag. Plotted as coloured squares are those sources that we have in common with \citet{2000AJ....119.2780A}, including those we rejected as foreground or due to the quality of their photometry. Sources classified as AGB are in red, those classified as RGB are in green, those classified as foreground are in yellow and those sources that were rejected by the two-flag quality criteria are in cyan. The dashed horizontal line marks the position of the TRGB and the vertical lines mark the ($J-K$) selection criteria applied in this work. Right: Colour histograms, using $0.02$~mag bins, showing the ($J-H$) and ($J-K$) colour distributions of those sources we have in common with \citet{2000AJ....119.2780A}, all of which they have classified as C-type according to the $CN-TiO$ method.}} 
\label{Albe}
\end{figure*}

In order to be
considered a reliable match, the sources in our catalogue and those in
the catalogue of \citet{2000AJ....119.2780A} must lie within $0.8$''
of each other. We found $145$ sources in common with the catalogue of
\citet{2000AJ....119.2780A}. We have classified $105$ ($72\%$) of
them as C-type stars. Of the remaining $40$ stars, we have classified
$25$ as M-type AGB stars, $2$ as RGB stars, $5$ were removed from our
sample as foreground stars and $8$ were rejected which
we have in common with \citeauthor{2000AJ....119.2780A}. We plot the sources common to both samples in Fig. \ref{Albe}, the strong
diagonal branch of the C-type giant population is clearly visible,
extending blueward of the ($J-K$)~$= 1.15$~mag colour separation we
have used for the $JHK$ classification the C- and M-type stars. This
is expected, as we foreshadowed in Sect. \ref{colsel}; the wide
colour distribution of C-type AGB stars means that while it is
possible to impose a red limit for the classification of M-type AGB
stars, it is far more difficult to impose a blue limit on the C-type
population. \citet{2012A&A...537A.108K}, \citet{2009A&A...493.1075B}
and \citet{2013..InPreparation...S} have all shown that there is an
overlap in ($J-K$) between the C- and the M-type stars and that no one
criterion or set of colour criteria can completely separate the
two populations. However, the ($J-K$) colour
distribution in the top right panel of Fig. \ref{Albe} shows, if we
assume the classifications of \citet{2000AJ....119.2780A} are correct,
the majority of the C-type sources lie redward of ($J-K$)~$=
1.15$. During their spectroscopic work on AGB stars in NGC 6822
selected using the $CN-TiO$ method, \citet{2013..InPreparation...S}
found that $JHK$ misclassifications in the $CN-TiO$ sample were most
likely to occur at bluer colours, where the $CN-TiO$ method does not
distinguish hotter C-type stars well
\citep{2006pnbm.conf..108G,2006A&A...456..905D,2002AJ....123..832L}. In
order to confirm any classification made using $JHK$ or $CN-TiO$
photometry, spectral data are required. 

In the bottom right panel of
Fig. \ref{Albe} we show the ($J-H$) colour distribution of the sources
we have in common with \citeauthor{2000AJ....119.2780A} the dashed
line marks the position of the ($J-H$) cut used in this work to
distinguish between foreground and genuine IC~1613 sources. The
majority of the sources classified as C-type by
\citeauthor{2000AJ....119.2780A} have ($J-H$)~$> 0.64$~mag; with only
$7$ sources having colours bluer than this limit, the ($J-H$)
distribution of these sources strongly supports our choice of foreground
colour criteria.

If it is assumed that the classifications of
\citet{2000AJ....119.2780A} are correct in all cases, then it suggests
that we may have misclassified up to $19\%$ of the genuine C-type
stars in IC~1613 as M-type. This would increase the C/M ratio within
the central $4.5$~kpc to $0.71 \pm 0.05$ and reduce the derived
metallicity to [Fe/H]~$= -1.32 \pm 0.07$~dex. However, the work of
\citet{2013..InPreparation...S} has shown that a similar catalogue of
$CN-TiO$ classified C-type stars in the dwarf galaxy NGC 6822
\citep{2002AJ....123..832L} has overestimated the number of C-type
stars by $\sim 7\%$. If we assumed the same level of error
in the classifications of \citet{2000AJ....119.2780A} then the
potential number of C-type stars misclassified as M-type in our own
AGB sample is reduced to $10\%$. In this case the C/M ratio in the central
$4.5$~kpc becomes $0.61 \pm 0.04$ and the metallicity [Fe/H]~$= -1.29
\pm 0.07$~dex.

\subsection{$J-K$ blue limit sensitivity}
\label{senJK}
In Sect. \ref{blue}, we established a blue limit at ($J-K$)~$= 0.75$~mag to exclude K-type sources from the M-type giant population. In addition to the small leakage of foreground dwarfs into the M-type giant box, evident in Fig. \ref{JKfg}, there is a suggestion in Fig. \ref{Albe} that a noticeable number of K-type giants in IC~1613 have also leaked into the M-type giant selection box. They are suggested by the presence of giants with $K < 17$~mag and ($J-K$)~$< 0.9$~mag, like those seen in Fig. \ref{FGspace}. If these $45$ stars are not M-type giants we would need to reduce the M-type giant count in Sect. \ref{c/m} from $552$ to $507$. This would increase the C/M ratio to $0.57$, implying a slightly lower metallicity of [Fe/H]~$= -1.28 \pm 0.07$~dex

\subsection{$J-H$ sensitivity}
\label{senJH}
In Sect. \ref{fore} we established a colour cut at ($J-H$)~$= 0.64$~mag to remove the majority of the foreground contamination in our
sample. However, it is probable that some genuine IC~1613 giants
(mainly M-type giants) were also removed when this colour criterion
was applied. We therefore consider if a bluer cut in ($J-H$) would be
more suitable and would preserve more M-type giants (but also admit more K-type giant contaminants). In the right hand
panel of Fig. \ref{CCDfg} we compared the ($J-H$) colour distributions
of a sample of foreground sources and a sample dominated by genuine
IC~1613 sources. The overlap between the two
samples is clear, but the foreground dwarf sample declines suddenly at
($J-H$)~$= 0.58$~mag and again at $0.64$~mag before tailing off at
redder colours, while the IC~1613 sample covers a broader range of
colours and peaks between ($J-H$)~$= 0.7-0.9$~mag. A colour cut at
$0.64$~mag was chosen to remove the maximum number of dwarf sources
from our sample, however from Fig. \ref{CCDfg} we see that a
significant number of genuine IC~1613 sources were also eliminated at
($J-H$)~$< 0.64$~mag, and we now consider the impact of using a bluer
($J-H$) cut at $0.58$~mag, on our derived C/M ratio. 

Within the
$4.5$~kpc radial limit applied in Sect. \ref{stellden}, we found $291$
C-type and $552$ M-type AGB stars. With the application of the bluer
($J-H$) criteria, this becomes $295$ C-type and $688$ M-type stars,
from which we calculate a C/M ratio of $0.43 \pm 0.03$ and an [Fe/H]
abundance of $ -1.22 \pm 0.08$~dex. As we would expect based on the
($J-H$) distribution of the C-type stars in Fig. \ref{Albe}, the bluer
colour cut has far more impact on the number of M-type giants than on
the number of C-type giants. This is also reflected in the calculated
foreground corrections required when using the bluer ($J-H$) cut,
$\sim 0.31$ C-type and $\sim 1.82$ M-type stars per kpc$^{2}$. After
making a statistical correction to account for the remaining
foreground interlopers the C/M ratio is $0.48 \pm 0.04$ and [Fe/H]~$=
-1.24 \pm 0.07$~dex. The [Fe/H] abundance derived using the bluer
($J-H$) cut is in good agreement with the value we derive in
Sect. \ref{c/m} ($-1.26 \pm 0.07$~dex), and therefore the change in
colour cut appears to have had little impact. Using the bluer ($J-H$)
cut means that the number of contaminants remaining in the central
grid region increases from $21$ to $31$ ($\sim 1.27$ stars
per~kpc$^{2}$) but the estimated foreground contamination in the AGB
samples is still $< 1\%$, although this will increase with radial
distance. We prefer to use a foreground colour cut at 
($J-H$)~$= 0.64$~mag in order to maintain the purest sample of giants.

\subsection{A comparison with NGC 6822}
\label{ngc6822}
During the $JHK$ photometric and subsequent spectroscopic studies of
NGC~6822 \citep{2012A&A...540A.135S,2013..InPreparation...S} the
following criteria were established for the selection C- and M-type
AGB stars: $K_{TRGB} = 17.41 \pm 0.11$~mag, ($J-H$)~$\geq 0.76$~mag,
($J-K$)~$\geq 0.93$~mag and a colour  of ($J-K$)~$= 1.17$~mag was used
to separate the C- and the M-type sources. It is interesting to note
that the absolute TRGB magnitude for NGC~6822 (M$_{K} = -6.04$~mag) is
almost the same as that of IC~1613 (M$_{K} = -6.15$~mag), and that the
($J-K$) colour separation is very similar for both galaxies. Based on the
C/M ratio of NGC~6822, the global metallicity of that AGB population
was found to be [Fe/H]~$-1.38 \pm 0.06$~dex (following our
spectroscopic study). 

Given the agreement between the calculate
metallicities for NGC~6822 and IC~1613, variations in the TRGB
magnitude of each galaxy may be used to compare the relative ages of
the AGB populations in the two galaxies. The close agreement between
the absolute magnitude of the TRGB in IC~1613 and in NGC~6822 suggests
that the two AGB populations are of similar age. Both IC~1613 and
NGC~6822 are thought to have undergone almost continuous star
formation until now
\citep{1999AJ....118.1657C,2003ApJ...588L..85C,2006AJ....131..343D},
although there has been no detailed study of the age of the AGB
population in either galaxy. However, based on an adopted abundance of
$Z = 0.001$, which is in good agreement with the metallicity we
derive, \citet{2009A&A...493.1075B} estimate that the majority of the
C-type star population in IC~1613 have ages between $700$~Myr and
$2.5$~Gyr, while \citet{2012A&A...537A.108K} suggest that the AGB
population of NGC~6822 have ages between $1.0 - 1.5$ Gyr based on the
fitting of isochrones with metallicities between [Fe/H]~$= -0.7$ to $-
1.3$~dex. Despite the wide range of metallicities used by
\citet{2012A&A...537A.108K}, the ages derived by
\citet{2009A&A...493.1075B} and \citet{2012A&A...537A.108K} also
suggest that the AGB populations of the respective galaxies are of a
similar age.

\section{Conclusions}
\label{cons}
We have used $JHK$ photometry of an area of $\sim 0.80$ deg$^{2}$
centered on the Irr dwarf galaxy IC~1613 to isolate the AGB population
of that galaxy, and have derived a global iron abundance based on the
C/M ratio of that population. The magnitude of the TRGB has been
determined and variations in that magnitude have been investigated as
a function of distance from the galactic centre and as a function of
azimuthal angle. The C/M ratio and the calculated [Fe/H] values have
been investigated as functions of the same parameters.

\noindent Below we summarise our main conclusions.

\begin{enumerate}
\item The MW foreground population, which for our purposes constitutes
  contamination of the desired photometric sample of IC~1613 giants,
  has been removed using a colour cut at ($J-H$)~$= 0.64$~mag. This
  cut has effectively removed the low-level foreground dwarf
  contamination in the direction of IC~1613, although some
  contamination is expected to remain due to the imperfect nature of
  the colour selection; this is estimated to be $\sim 0.80$ stars
  per~kpc$^{2}$.

\item The TRGB magnitude has been measured to be $K_{0} = 18.28 \pm
  0.15$~mag using the Sobel edge-detection algorithm. This value has
  been used to select AGB stars from the IC~1613 population. Further
  measurements of the TRGB at different galactocentric distances and
  as a function of azimuthal angle have not found any trend in the
  TRGB magnitude with either parameter.   

\item Colour selection criteria of $ 0.75 \leq$ ($J-K$)~$< 1.15$~mag
  and ($J-K$)~$\geq 1.15$~mag have been used to classify M- and C-type
  AGB stars respectively. The error on the colour boundary at ($J-K$)
  $= 1.15$~mag is estimated to be $\sim 0.05$~mag. This colour value
  translates into ($J-K$)$_{2MASS} = 1.22$~mag for photometric system
  of the 2MASS point source catalogue. A blue limit of ($J-K$)~$=
  0.75$~mag has been applied to the M-type star population in order to
  exclude K-type giant sources and provide a purer M-type sample for
  our determination of the C/M ratio.  

\item Using the measured stellar density in annuli at intervals of
  $0.5$~kpc in the de-projected plane of the galaxy, the density
  profile of the AGB population has been plotted and traced out to a
  distance of $6$~kpc from the centre of the galaxy. Based on this
  result we have used only those sources within $4.5$~kpc for the
  determination of the C/M ratio and the [Fe/H] abundance within
  IC~1613.   

\item A C/M ratio of $0.52 \pm 0.04$ has been derived within the
  $4.5$~kpc radial limit we have imposed. From this value we have
  calculated an [Fe/H] abundance of $-1.26 \pm 0.07$~dex, \textbf{using C/M~vs.~[Fe/H] relation presented by \citet{2009A&A...506.1137C}. Although we have not been able to incorporate the affects of the stellar age distribution in our calculations, the value we derive is consistent with other population metallicity measurements in the literature and we see very little variation in iron abundance cross the galaxy (Sect. \ref{c/m}). For IC~1613 this first-order approximation of the C/M~vs.~[Fe/H] relation appears to provide a good estimate of the global galactic metallicity.}

\item The ($J-H$) and ($J-K$) colour distribution of the C-type stars
  population has been examined in more detail based on a sample of our
  photometric sources in common with
  \citet{2000AJ....119.2780A}. Based on our $JHK$ photometry for $145$
  candidate C-type sources presented by
  \citet{2000AJ....119.2780A}, we find that only $7$ have ($J-H$)
  $<0.64$~mag. The ($J-K$) distribution of these sources shows that
  C-type stars are present on both sides of our colour boundary for
  the classification of the AGB sources, but that the majority have
  colours ($J-K$)~$\geq 1.15$~mag.  

\item Based on a comparison with the work of
  \citet{2000AJ....119.2780A} who relied on the $CN-TiO$ method to
  classify C-type AGB stars in IC~1613, and the work of
  \citet{2013..InPreparation...S} who provide an estimate of the error
  associated with that method, it was estimated that up to $10\%$ of
  the C-type population in our sample may have been misclassified as M-type
  stars. Correcting for this potential misclassification in our
  photometric sample results in a C/M ratio of $0.61 \pm 0.04$ and [Fe/H]~$= -1.29 \pm 0.07$~dex.     

\item In order to preserve the maximum number IC~1613 sources a bluer
  ($J-H$) cut was also examined. Using ($J-H$)~$= 0.58$~mag, instead
  of $0.64$, to remove the dwarf foreground sources results a C/M
  ratio of $0.48 \pm 0.04$ and an iron abundance of [Fe/H]~$= -1.24
  \pm 0.07$~dex. The impact of the change in colour cut is small and
  the derived iron abundances for each cut are in good agreement. It
  was decided to use the redder ($J-H$) cut in order to preserve the
  purest sample of giant stars.     
\end{enumerate}

\begin{acknowledgements}
We would like to take the opportunity to thank the referee for his/her detailed reading of this paper and the useful comments provided.
\end{acknowledgements}

\bibliographystyle{aa}
\bibliography{/home/lizmonster/Current/LizIC1613p1}

\end{document}